%% file: main.tex
\documentclass[12pt]{article}
\usepackage{a4}
\usepackage{epsfig}
\def \sle  {\ensuremath{ \tilde{\ell}               }}
\def \sleR {\ensuremath{ \tilde{\ell}_\mathrm{R}    }}

\def \selR {\ensuremath{ \mathrm{\tilde{e}_R}       }}
\def \selL {\ensuremath{ \mathrm{\tilde{e}_L}       }}

\def \neu   {\ensuremath{ \mathrm{\chi^0}           }}
\def \neue  {\ensuremath{ \mathrm{\chi^0_1}         }}
\def \neuz  {\ensuremath{ \mathrm{\chi^0_2}         }}
\def \neud  {\ensuremath{ \mathrm{\chi^0_3}         }}

\def \phot {\ensuremath{ \gamma                     }}
\def \el     {\ensuremath{\mathrm{e}                }}
\def \lep    {\ensuremath{ \ell                     }}

\def \mvis   {\ensuremath{ M_{\rm{vis}}            }}

\def \msnu   {\ensuremath{ m_{\tilde{\nu}}         }}
\def \msle   {\ensuremath{ m_{\sle}                }}

\def \mselR  {\ensuremath{ m_{\selR}               }}
\def \mselL  {\ensuremath{ m_{\selL}               }}

\def \mtop   {\ensuremath{ m_{\rm top}             }}
\def \mh     {\ensuremath{ m_{\rm h}               }}
\def \ma     {\ensuremath{ m_{\rm A}               }}
\def \mchi   {\ensuremath{ m_{\neu}                }}
\def \mneue  {\ensuremath{ m_{\neue}               }}

\newcommand{\grsim}{\,\raisebox{-0.6ex}{$\stackrel{\textstyle>}{\textstyle\sim}$}\,}
\def \degg {\ensuremath{^{\circ} }}

\def \geqq {\ensuremath{ \geq }}

\def \pro    {\ensuremath{ \%                                   }}  
\def \tanb   {\ensuremath{ \tan{\beta}                          }}
\def \deltm  {\ensuremath{ \Delta M                             }}
\def \deltme {\ensuremath{ \Delta M_{\neuz\neue}                }}
\def \deltmz {\ensuremath{ \Delta M_{\sleR\neuz}                }}
\def \rts    {\ensuremath{ \sqrt{s}                             }}

\def \pe     {\ensuremath{ p_{\mathrm 1}                        }}

\def \pte    {\ensuremath{ p_{\mathrm T1}                       }}
\def \phiaco {\ensuremath{ \Phi_{\mathrm{aco}}                  }}
\def \mgev   {\ensuremath{ \, \mathrm{GeV}/\it{c}^{\mathrm{2}} }}
\def \pgev   {\ensuremath{ \, \mathrm{GeV}/\it{c}              }}
\def \egev   {\ensuremath{ \, \mathrm{GeV}                      }}

\def \nbarnf {\ensuremath{ \bar{N}_{95}                         }}
\def \epem   {\ensuremath{\mathrm{e^+e^-}                       }}
\def \bepem  {\ensuremath{e^+e^-                                }}
\def \ncand  {\ensuremath{ N_{\rm obs}                          }}
\def \nbkg   {\ensuremath{ N_{\rm exp}                          }}

\def \tanb   {\ensuremath{ \tan{\beta}         }}
\def \cotb   {\ensuremath{ \cot{\beta}         }}
\def \Mu     {\ensuremath{ \mu                 }}

\def \mn     {\ensuremath{ m_0                 }}
\def \mez    {\ensuremath{ m_{1/2}             }}

\def \Af     {\ensuremath{  A_{\rm{f}}         }}
\def \An     {\ensuremath{  A_0                }}
\def \astop  {\ensuremath{ A_{\rm t}           }}

\begin{document}
\parskip 0.2cm plus 0.05cm
\font\eightrm=cmr8
\font\ninerm=cmr9
\date{}
\title{ \null\vspace{1cm}
Absolute Lower Limits \\on the Masses of Selectrons and Sneutrinos\\
in the MSSM\\
\vspace{1cm}}
\author{The ALEPH Collaboration$^*)$}

\maketitle

\begin{picture}(160,1)
\put(0,115){\rm ORGANISATION EUROP\'EENNE POUR LA RECHERCHE NUCL\'EAIRE
(CERN)}
\put(30,110){\rm Laboratoire Europ\'een pour la Physique des Particules}
\put(125,90){\parbox[t]{45mm}{\tt CERN-EP/2002-055}}
\put(125,84){\parbox[t]{45mm}{\tt 15th July 2002}}
\end{picture}

\vspace{.2cm}
\begin{abstract}
\vspace{.2cm}
The results of searches for selectrons, charginos and neutralinos
performed with the data collected by the ALEPH detector at LEP at
centre-of-mass energies up to 209\,GeV are interpreted in the framework of
the Minimal Supersymmetric extension of the Standard Model with R-parity
conservation. Under the assumptions of gaugino and sfermion mass
unification and no sfermion mixing,
an absolute lower limit of 73\,GeV/$c^2$ is set on the mass of the lighter
selectron $\tilde{\rm e}_{\rm R}$ at the 95\% confidence level. Similarly,
limits on the masses of the heavier selectron $\tilde{\rm e}_{\rm L}$ and
of the sneutrino $\tilde{\nu}_{\rm e}$ are set at 107 and 84\,GeV/$c^2$,
respectively. Additional constraints are derived from the results of the
searches for Higgs bosons. The results are also interpreted in the framework
of minimal supergravity.
\end{abstract}

\vfill
\centerline{\it Submitted to Physics Letters B}
\vskip .5cm
\noindent
--------------------------------------------\hfil\break
{\ninerm $^*)$ See next pages for the list of authors}

\eject

\input{authb}

\pagenumbering{arabic}


\section{Introduction}\label{sec:intro}

Supersymmetry~\cite{mssm} predicts the existence of a supersymmetric partner for
each Standard Model particle chirality state. 
In this letter, the results of standard searches for sleptons~($\tilde\ell$) 
and charginos ($\chi^\pm$) in \epem\, collisions,
already reported by ALEPH in \linebreak Refs.~\cite{slep1,char1,char2}, are combined with those of selectron 
and neutralino searches specifically developed for final states not 
considered in the former analyses. 
The results of these searches allow an absolute lower
limit to be set on the selectron and sneutrino mass.

The theoretical framework is the Minimal Supersymmetric extension of the
Standard Model (MSSM)~\cite{mssm}, with R-parity conservation and the
assumption that the lightest supersymmetric particle (LSP) is the lightest
neutralino $\chi_1^0$.
The notations and conventions of Ref.~\cite{char1}
are used for the MSSM parameters. The interpretation of the results
in terms of mass limits is done under the assumption of gaugino and
sfermion mass unification to common gaugino and scalar masses, $m_{1/2}$
and $m_0$, at the GUT scale. 
 At the electroweak scale, gaugino
masses are determined at tree level by $m_{1/2}$, the Higgs mass term $\mu$ and
$\tan\beta$, the ratio of vacuum expectation values of the two Higgs
doublets, assumed to be greater than 1 as is usual in the MSSM.

Charged and neutral slepton masses are expressed as indicated in
Eqs.~(\ref{eq:selectronR}--\ref{eq:sneutrino}), from which it can be seen that
the supersymmetric partner of the right-handed electron 
\selR\, is the lighter of the two selectrons.
\begin{eqnarray}
\label{eq:selectronR}
m_{\tilde\ell_{\rm R}}^2 & = & m_0^2 + 0.15\,m_{1/2}^2 -
m_{\rm Z}^2\,\cos 2\beta \sin^2\theta_{\rm W} \\
\label{eq:selectronL}
m_{\tilde\ell_{\rm L}}^2 & = & m_0^2 + 0.52\,m_{1/2}^2 -
{m_{\rm Z}^2 \over 2} \cos 2\beta \left ( 1 - 2\sin^2\theta_{\rm W}\right)
\\
\label{eq:sneutrino}
m_{\tilde\nu}^2 & = & m_0^2 + 0.52\,m_{1/2}^2 +
{m_{\rm Z}^2 \over 2} \cos 2\beta
\end{eqnarray}
Mixing effects, proportional to the mass of the Standard Model partner,
 are expected to be small for selectrons, and are therefore neglected throughout. 
For the results obtained in the MSSM, the mixing is set to zero for all sfermions
by enforcing 
the parameters \Af\, to their no-mixing values, $\Af=\Mu\tanb$ or 
$\Mu\cotb$ for down-type and up-type sfermions, respectively. 

The results of Higgs boson searches~\cite{higgs} are exploited to further
constrain the selectron and sneutrino masses at small $\tan\beta$ values,
for any values of the pseudoscalar neutral Higgs boson mass $m_{\rm A}$ and
of \astop, the trilinear coupling in the stop sector.

Tighter limits are also set in the framework of a highly constrained MSSM
version known as minimal supergravity (mSUGRA)~\cite{mssm}. In this model, 
$m_{\rm A}$ also derives from the common scalar mass $m_0$ at the GUT scale, 
the value $\vert \mu \vert$ is predicted from dynamical electroweak symmetry 
breaking, and the trilinear coupling  at the GUT scale, $A_0$, is common to all 
sfermions.
In this letter, $A_0 = 0$ is assumed.

The data used in the analyses entering the present combination were
collected with the ALEPH detector at LEP, at centre-of-mass energies ranging from 183 to
209\,GeV. The corresponding integrated luminosities are given in
Table~\ref{tab:lumi}. 
The results of the dedicated searches for selectrons
and neutralinos in the data collected
in the year 2000 are reported in this letter. These selections address
\begin{itemize}
\item[{\it (i)}] the $\tilde{\rm e}_{\rm R} \tilde {\rm e}_{\rm L}$
production to investigate, as described in Ref.~\cite{slep2}, small mass differences
between $\tilde{\rm e}_{\rm R}$ and $\chi_1^0$, for which the selections
of Ref.~\cite{slep1} become ineffective;
\item[{\it (ii)}]
the $\chi^0_1 \chi^0_3$ production with a subsequent neutralino decay into
slepton $\chi^0_3 \to \tilde\ell_{\rm R} \ell$, to cover specific regions of
the MSSM parameter space in which none of the selections of 
Refs.~\cite{slep1,char1,char2,slep2} constrain the mass of the selectron. 
\end{itemize}

This letter is organized as follows. In Section~2, the search strategy
towards an absolute lower limit on the selectron and sneutrino masses 
is explained.
The ALEPH detector is briefly described in Section~3.
The selections developed for the two specific final states mentioned 
above are presented in Section~4, and the interpretation of their results,
combined with those of previous analyses, is given in Section~5. 

All limits reported in this letter are at the 95\% confidence level.

\begin{table}[htbp]
\begin{center}
\caption{\footnotesize Integrated luminosities collected between 1997 and
2000 and average centre-of-mass energies.
\label{tab:lumi}}
\vspace{3mm}
\begin{tabular}{|l|c|c|c|c|c|c|c|c|c|} \hline
Year & 1997 & 1998 & \multicolumn{4}{|c|}{1999} & \multicolumn{3}{|c|}{2000}\\ 
\hline\hline
$\sqrt{s}$ (GeV) & 182.7 & 188.6 & 191.6 & 195.5
                 & 199.5 & 201.6 & 205.2 & 206.6 & 208.0 \\ 
\hline
${\cal L}$ (${\rm pb}^{-1}$)
                 &  56.8 & 173.6 & 28.9  &  79.8
                 &  86.2 &  42.0 & 75.3  & 122.6 & 9.4 \\ 
\hline\hline
\end{tabular}
\end{center}
\end{table}

\section{Search strategy \label{s:strategy}}

An absolute lower limit on the lighter selectron mass can be derived 
by a scan of the MSSM parameters, $m_0$, $m_{1/2}$, $\tanb$ and
$\mu$. In the absence of sfermion mixing, these parameters suffice to 
determine the masses and couplings of gauginos and sleptons at tree level, 
and therefore the relevant production
cross sections and decay branching fractions. The
scan is performed for $\tan\beta$
between 1 and 50, and  $\mu$ between $-10$ and $\rm +10\,\rm TeV/c^2$.
For \mn\, and $m_{1/2}$, the explored range is limited to values smaller than from 
 100 to 200\,\mgev\, to keep the selectron masses
below the LEP kinematic limit. The scan of $m_{1/2}$ is further bounded
from below by the absolute lower limit on the mass of the LSP, 
set at 37\,\mgev\, in Ref.~\cite{char1}
under the same hypotheses as those used in this letter.

For large mass differences \deltm\, ($\deltm \grsim 10\,\mgev\,$)
between the lighter selectron and the 
lightest neutralino, the standard \selR\selR\, searches~\cite{slep1}
apply as long as  selectrons decay predominantly into \el\neue, 
and allow selectron masses to be excluded beyond 90\,\mgev.
The efficiency of this selection decreases with \deltm. 
Indeed, for small \deltm\, values (below~4\,~\mgev), the searches can barely improve 
on the limit obtained from the Z width measurement at LEP\,1~\cite{lep1}.

However, this case can be covered as described in Ref.~\cite{slep2} by a
search for the \selR\selL\, associated production, 
with the subsequent decay of both sparticles into e\neue. In this final state, at
least one energetic electron stems from the decay of the heavier selectron.  
For \deltm\, values in excess of
$\sim 2\mgev$, the additional low-momentum electron remains visible. 
The standard selectron searches are therefore efficient at selecting the   
\selR\selL\, production by merely adapting
the electron momentum sliding cuts as a function of the lighter selectron  
and neutralino masses.
For very small \deltm\, values, the low momentum electron is not reconstructed, 
such that the final state consists of a single electron and  missing energy.  
The results of the search for this topology,
investigated in Ref.~\cite{slep2} and updated at centre-of-mass energies up
to 202\,GeV in Ref.~\cite{char1}, are reported here at the highest energies
produced by LEP\,2 in the year 2000.

In large parts of the parameter space, the lower limit on the selectron mass
is set by using a combination of the above-mentioned selectron searches.  
Problems occur in certain regions characterized by small values of 
\tanb, $|\Mu|$ and \mn, which lead to light \neuz\, with a high photino field 
content. Predominant selectron cascade decays via the \neuz\, yield final states not
selected by the standard selectron searches. In this case, the charginos are
also light such that the region is in general excluded by 
chargino searches. This coverage vanishes in the so called 
{\it corridor}~\cite{char1}, a subset of model parameters
for which the chargino and the sneutrino are degenerate in mass.
In the corridor, the final state arising from the chargino two-body (2B) 
decay into $\tilde{\nu}\ell$,  
dominant over the three-body (3B) decay into 
$\neue\rm f\bar{f}$, is in practice invisible.

In this case, though, the $\chi_3^0$ is
light enough for  the \neue\neud\, production cross section
 with a $\chi^0_3$ decay into $\tilde\ell \ell$ to be  sufficient
to cover this peculiar region. Searches for six different final states,
according to the slepton decay (direct or cascade) and flavour (e or $\mu$),
were developed to address the associated neutralino production, and their
results are reported here.

The low-$\tan\beta$ region is also covered by the result of the
searches for the lighter neutral scalar Higgs boson h~\cite{higgs}, as described in
Ref.~\cite{char1}. However, because the lower limit on \tanb\, varies rapidly
with the top quark mass through radiative corrections to
\mh, the uncertainty on \mtop\, renders this indirect limit
less robust than that obtained with the direct selectron searches.

\section{The ALEPH detector}

A thorough description of the ALEPH detector and its performance as well as of the standard 
reconstruction and analysis algorithms can be found in Refs.~\cite{detector,performance}. 
Only a brief summary is given here. 

The trajectories of charged particles are 
measured by a silicon vertex detector, a cylindrical multi-wire drift chamber and 
a large time projection chamber (TPC). Charged
particle trajectories are called {\it good tracks} if they are reconstructed
with at least four space points in the TPC, a transverse momentum in excess
of 200\,MeV/$c$, a polar angle with respect to the beam such that
$\vert\cos\theta\vert < 0.95$, and if they originate from within a cylinder of
length 20\,cm and radius 2\,cm coaxial with the beam and centred at the
nominal interaction point. In addition, good tracks must not be compatible
with arising from a photon conversion to \epem\,
identified as pairs of oppositely-charged particles satisfying stringent
conditions on their distance of closest approach and their invariant mass.

The tracking devices are immersed in an axial magnetic field of 1.5\,T, 
provided by a superconducting solenoidal coil and surrounded by a highly 
segmented electromagnetic calorimeter (ECAL).
The ECAL is used to identify electrons and photons by the
characteristic longitudinal and transverse developments of the associated
showers, and is supplemented for low momentum electrons by the measurement
in the TPC of the specific energy loss by ionization.

The iron return yoke is instrumented with streamer tubes as a hadron calorimeter (HCAL).
It provides a measurement of the hadronic energy and,
together with external muon chambers, efficient identification of muons
by their characteristic penetration pattern.
Luminosity monitors extend 
the calorimeter coverage down to 34\,mrad.

Global event quantities such as total energy, transverse momentum or
missing energy, are determined from an energy-flow algorithm which combines
all the above measurements into charged particles (electrons, muons, charged
hadrons), photons and neutral hadrons, which are the basic objects used in 
the selections presented in this letter.


\section{Event Selection}
\label{s:selec}

The selection criteria described below were optimized with the
\nbarnf\, prescription \cite{nbar95} which consists in minimizing the upper
limit on the signal cross section expected in the absence of signal
processes. To do so, the selections were applied to fully simulated standard
model background samples, generated as in Ref.~\cite{slep1} for $\epem \to {\rm f\bar f}$,
WW, ZZ, Zee, We$\nu$,
Z$\nu\bar\nu$, and for \phot\phot\, interactions. The simulation of the
associated \selR\selL\, and \neue\neud\,
production was performed with SUSYGEN~\cite{susygen}.

\subsection{Update of the search for associated \selR\selL\, production \label{s:leftright} }

The selection of single-electron final states, described in detail in Ref.~\cite{slep2}, 
was applied to the highest
centre-of-mass energy data, with the kinematic cuts appropriately rescaled.
The numbers of candidate events observed and background events expected 
are given in Table~\ref{t:singlee}, together with the results of previous 
years~\cite{char1,slep2,slep3}.

\begin{table}[hbt]
\renewcommand{\arraystretch}{1.2}
\caption{\footnotesize Numbers of candidate events observed (\ncand) and background events 
expected (\nbkg) for the
single-electron selection. \label{t:singlee} }
\begin{center}
\begin{tabular}{|c||c|c|}
\hline \hline
Energy (\egev) & \nbkg  &\ncand  \\
\hline \hline
182.7          &   6.6   & 5 \\ 
\hline
188.6          &  13.8   & 8 \\
\hline
191.6          &   2.7   & 2 \\
195.5          &   7.5   & 9 \\
199.5          &   8.2   & 9 \\
201.6          &   4.2   & 2 \\         
\hline
205.2          &   7.7   & 5 \\
206.6          &  13.8   & 7 \\
208.0          &   1.0   & 0 \\
\hline
Total            &  65.5   & 47 \\
\hline\hline
\end{tabular}
\end{center}

\end{table}

\begin{figure}[h]
\setlength{\unitlength}{1.0cm}
\begin{center}
\begin{picture}(12.0,7.0)
\put(-2.0,0.0){\epsfig{file=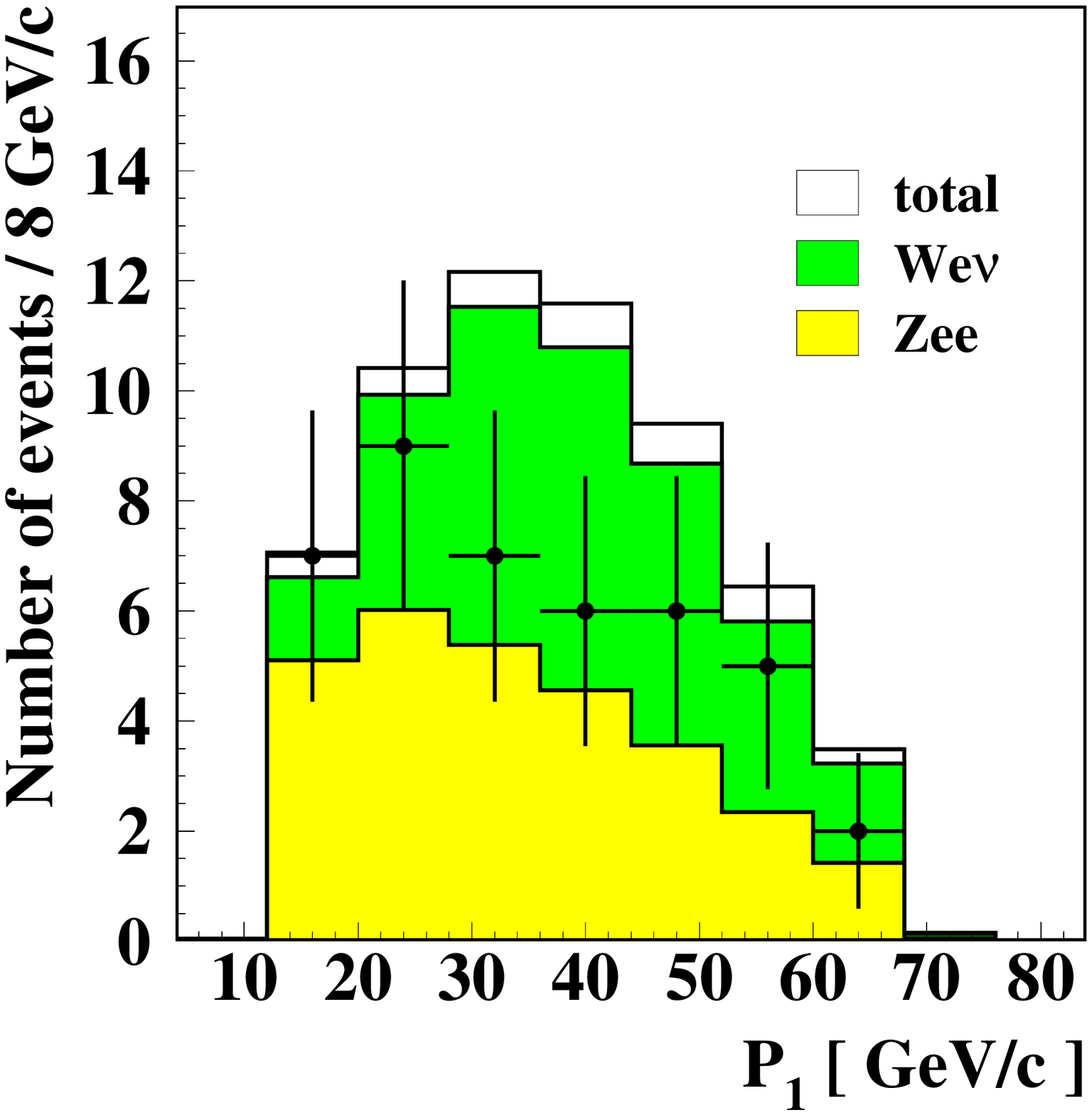,%
            height=8cm,width=8cm}}
\put(7.0,0.0){\epsfig{file=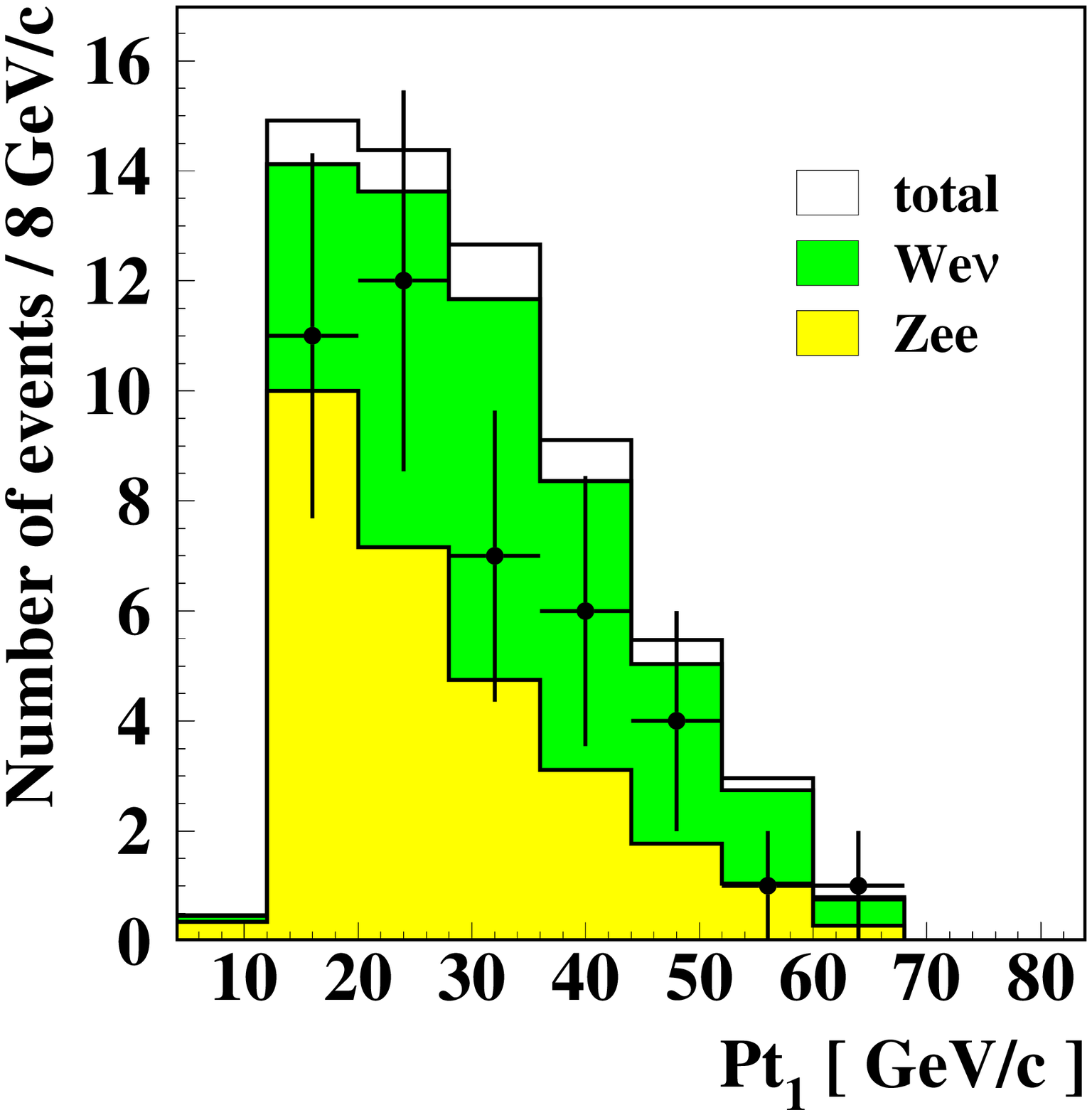,%
            height=8cm,width=8cm}}
\put(-0.3,6.7){(a)}
\put(8.7,6.7){(b)}
\end{picture}
\caption{\footnotesize Comparison between data (dots with error bars) and expected backgrounds 
(histograms) after  the single-electron selection cuts for centre-of-mass energies ranging from
189 to 209 GeV. Distributions of (a) the momentum $\rm P_1$ and (b) the transverse momentum 
$\rm Pt_1$ of the leading electron.
\label{f:PtDist}}
 \end{center}
 \end{figure}

The total number of events in the data is significantly smaller than
expected from standard background sources, dominated by the processes $\epem\to\rm We\nu$ 
and $\epem\to\rm Zee$.  
A study of this 2.2 standard deviation deficit
led to the conclusion that it is unlikely to be of systematic origin. In
particular, the distributions of all relevant kinematic quantities are in
qualitative agreement with those expected from the production of the 
$\rm We\nu $ and the Zee final states as is exemplified in  Fig.~\ref{f:PtDist} 
for the total and the transverse momentum of the leading electron at 
centre-of-mass energies in excess of  188.6~\egev. 
The total cross sections predicted for 
$\epem\to\rm We\nu$ by several generators (PYTHIA \cite{pythia},
GRACE4F \cite{grace4f}) show no difference beyond the 10\% level, which
would account for only a third of the effect. Notwithstanding the probable 
statistical origin of this deficit, it is conservatively taken
into account in deriving signal cross-section upper limits as is explained in
Section~\ref{s:sys}.

\subsection{Search for associated \neue\neud\, production    \label{s:neutr} }

The \neud\, decay modes considered in the present analysis are listed 
in Table~\ref{t:decay}. The \neud\, decay via selectrons or smuons leads
to final states with electrons, muons and possibly  photons.
The selection criteria used in these analyses are based on the acoplanar
lepton or single electron analyses.   
The selection cuts are optimized and adapted for each specific 
decay topology.
The number of good tracks in the detector depends on the decay mode and 
on the mass differences involved.

\begin{table}[bt]
\renewcommand{\arraystretch}{1.2}
\caption{\footnotesize Leptonic \neud\,  decay modes and final states.
 In the present analysis only decay modes via selectrons and smuons are considered.
\label{t:decay}}
\begin{center}
\begin{tabular}{|lccccccll|}
\hline
\hline 
\multicolumn{3}{|l}{decay mode}&&&&&&final state\\
\hline
\hline
A & \neud & $\to$ & \sleR & $+$           & \lep &     &         & one or two leptons   \\
  &   &               & \sleR & $\to$ & \neue   & $+$ & \lep &     \\

\hline
B & \neud & $\to$ & \sleR & $+$           & \lep &     &         & one to four leptons  \\
  &   &               & \sleR & $\to$ & \neuz   & $+$ & \lep &              \\
  &   & & &           & \neuz & $\to$ & \neue $+$ \lep\lep & \\

\hline
C & \neud & $\to$ & \sleR & $+$           & \lep &     &         & one or two leptons    \\
  &   &               & \sleR & $\to$ & \neuz   & $+$ & \lep & and a photon\\
  &   & & &           & \neuz & $\to$ & \neue $+$ $\gamma$ &     \\
\hline 
\hline
\end{tabular}
\end{center}
\end{table}

\begin{itemize}

\item In decay mode A,  
 one or two good tracks (electrons or muons) are expected.
The acoplanar lepton and the single electron searches are
therefore applied as preselections with no modification. The latter is
extended to also cover the single muon topology, by substituting muon
identification for electron identification.

\item In  decay mode B, the \neud\, decay yields a high momentum lepton
and typically three softer leptons. For large \deltme\, 
at least three good tracks (electrons or muons) are required,
and the acoplanar lepton search is applied as a preselection on the two
leading tracks. The single-electron or single-muon selections, modified to
accept up to three low momentum tracks, bring additional efficiency for
small \deltme\, and \deltmz\, values.

\item In decay mode C, one or two good tracks are expected,
accompanied with an energetic photon.
The acoplanar-lepton and the single-lepton selections are therefore applied,
modified by requiring a photon with an energy in excess of 5 GeV. Furthermore,
its angular separation and invariant mass with any good track must be larger than
10\degg\, and 2~\mgev, respectively.

\end{itemize}

Because of the specific kinematics of the final states arising from
\neue\neud,  other common requirements are applied. In particular,
the presence of two invisible $\chi^0_1$'s leads to large missing energy. To
reduce the dominant backgrounds from WW and ZZ processes, the visible mass
is therefore required to be smaller than 80\,GeV/$c^2$. Moreover, as the
leading lepton is expected to be more energetic in \neue\neud\,
production than in direct selectron production, the cuts on its momentum
$p_1$ and its transverse momentum $p_{T1}$ are significantly tightened. The
additional background from $\gamma\gamma$ processes selected with three or
four good tracks is efficiently reduced by these cuts. Finally, the
acoplanarity cut ($\Phi_{\rm aco}$) is relaxed in some cases to preserve a
reasonable signal efficiency. The basic selections and the
additional/modified criteria for each of the final states are summarized in
Table~\ref{t:x13selec}.

\begin{table}[bt]
\renewcommand{\arraystretch}{1.2}
\caption{\footnotesize Selection criteria for the three \neue\neud\, decay modes A, B and C 
for different final states. 
The numbers of additional good tracks with low momentum are indicated in brackets for each of
the final states.\label{t:x13selec}}
\begin{center}
\begin{tabular}{|ll|l|l|}
\hline \hline
& Signature & Basic Selection & Modified Cuts                                             \\ 
\hline \hline
A & 2 \lep    & Acoplanar Leptons & $\mvis < 80 \mgev$, $\pe > 6\pro\rts$         \\
  & 1(+1) \lep & Single Electron  &Selection extended  to single muons                  \\ 
\hline
B & 3-4 \lep  & Acoplanar Leptons & Three to four good tracks  \\
  &           &                   & Standard cuts on the two leading tracks             \\
  &           &                   & $\mvis < 80\mgev$ , $\pe > 8\pro\rts$             \\
  & 1(+3) \lep & Single Lepton (A) & $\pte > 10\pro\rts$, \\
  &           &                   & Energy of additional good tracks $< 2\pro\rts$\\
  &           &                   & In case of  \geqq\, 2 good tracks: $\mvis > 4\pro\rts$,\\
  &           &                   & $\phiaco < 170\degg$ \\
\hline
C & 2\lep+\phot &  Acoplanar Leptons & One isolated \phot\, (see text), $E_\phot > 5\egev$, \\
  &             &                    & $\pte > 8\pgev$, $\mvis < 80\mgev$    \\
  & 1(+1)\lep+\phot & Single Lepton (A) & One isolated \phot\, (see text), $E_\phot > 5\egev$,  \\
  &                 &                   & For one good track: $\pe < 46.5\pro\rts$, $\phiaco < 175\degg$  \\
\hline\hline
\end{tabular}
\end{center}

\end{table}

The numbers of candidate events observed (\ncand) and background events expected
(\nbkg) for the three analyses are given in Table~\ref{t:x13cand}.

The results of analysis A are similar to those described in 
Ref.~\cite{slep1} and in Section~\ref{s:leftright} for the slepton searches.
The selection of three to four leptons in analysis B yields only a 
small number of expected background and candidate events. The deficit
in the selection of 1(+3) leptons is correlated to that in the single-electron 
selection presented in Section~\ref{s:leftright}. 
In analysis C, the background expectation is strongly suppressed by the requirement 
of a high-energy photon in the
final state. 

Finally, sliding cuts on the momenta of the leading two leptons are applied 
as a function of \msle\, and \mchi.
These cuts are defined by the momentum ranges expected for the kinematics of the
decay chain involved.
 
\begin{table}[bt]
\renewcommand{\arraystretch}{1.2}
\caption{\footnotesize Numbers of candidate events observed (\ncand) and background 
 events expected (\nbkg) for the \neue\neud\, selections in the year 2000 data set.
 \label{t:x13cand} }
\begin{center}
\begin{tabular}{|ll||c|c|}
\hline \hline
& Signature & \nbkg & \ncand\\ 
\hline \hline
A & 2 e & 16.5 & 18\\
  & 2 \Mu     & 15.0 & 23\\
  & 1(+1) e  & 22.5 & 12\\ 
  & 1(+1) \Mu & 13.7 & 9 \\
\hline
B & 3-4 \lep   &  4.4 & 2 \\
  & 1(+3) \lep & 33.6 & 17 \\
\hline
C & 2\lep+\phot &  2.1 & 3 \\
  & 1(+1)\lep+\phot & 3.8 & 4\\
\hline\hline
\end{tabular}
\end{center}

\end{table}


\section{Results\label{s:results}}

\subsection{Systematic uncertainties and cross section upper limits\label{s:sys}}

The main systematic uncertainties on the background and signal predictions 
arise from the statistics of the simulated
samples and from the simulation of the lepton identification~\cite{slep1}.
Both the background and signal expectations are conservatively reduced by this
uncertainty. The predicted background contribution  from $\rm\epem\to We\nu$ 
is further reduced by 10 $\%$ to account for the theoretical uncertainty of 
the production cross section.  
A systematic correction of $-14\%$ is also applied on the signal
efficiencies to account for the effect of the
cut on the energy detected at small polar angle~\cite{slep1}.

The optimal combination of selections is
chosen according to the \nbarnf\, prescription for each set of MSSM parameters tested.
To derive cross section upper limits, the dominant background 
($\epem\to\rm WW$ for the selections based on the
acoplanar-lepton search, and
$\epem\to$ Zee and We$\nu$ for those based on the single-track
search) is subtracted with the method of Ref.~\cite{pdg}.


\subsection{Limit on \mselR \label{s:rresults}}

In the MSSM framework, the upper bounds on the cross section allow lower
limits to
be set on the lighter selectron mass with a scan of the four relevant
parameters, $m_0$, $m_{1/2}$, $\mu$ and $\tan\beta$, as described in
Section~\ref{s:strategy}.
These limits are presented here in the (\mn, \mez) plane.

For large $\tan\beta$ values (in excess of $\sim 7$), small differences
$\Delta M$ between the \selR\, and $\chi_1^0$ masses are only allowed
for large \neue\ masses. Any selectron mass below
92\,GeV/$c^2$ is therefore excluded by the standard \selR\selR\,
searches in this region of the parameter space.

\begin{figure}[h]
\setlength{\unitlength}{1.0cm}
\begin{center}
\begin{picture}(12.0,10.0)
\put(0.0,0.0){\epsfig{file=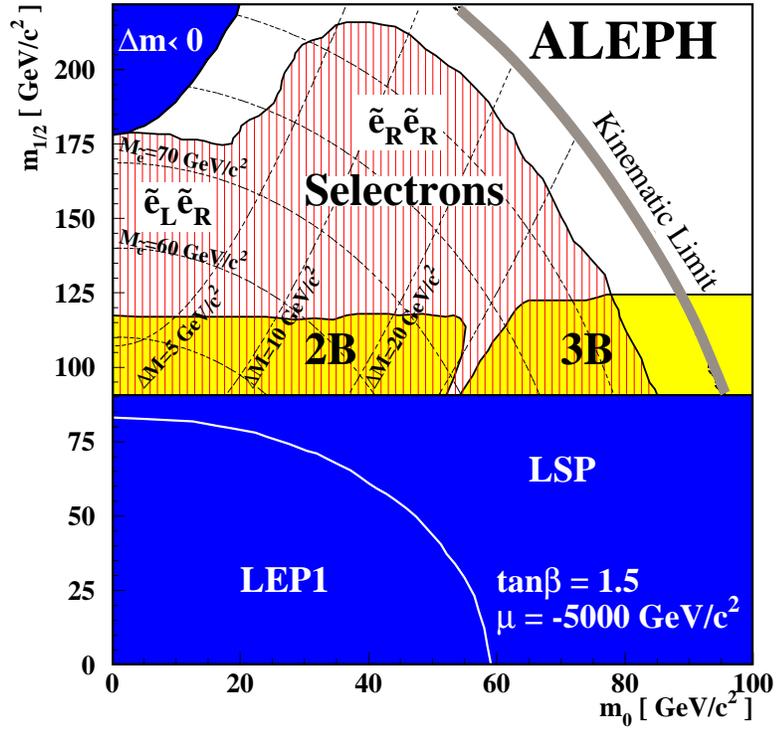,%
            height=11cm,width=11cm}}
\end{picture}
\caption{\footnotesize 
Regions excluded in the (\mn, \mez) plane by direct searches for selectrons and charginos  
for $\tanb = 1.5$ and $\mu =-5000\mgev$.
The dark-shaded regions are theoretically forbidden ($\deltm< 0$) or  
excluded by LEP1 \cite{lep1} or the LSP limit \cite{char1}.
The light-shaded regions are excluded by direct 
searches for chargino two-body (2B) and three-body (3B) decays \cite{char1, char2}. 
The hatched region is excluded by the selectron search.
Lines of constant \deltm~values  and of constant \mselR\,values  are also shown.
The thick line indicates the kinematic limit for \selR\selR\, production.}
\label{f:m0m2}
 \end{center}
 \end{figure}

For smaller $\tan\beta$ values,
small $\Delta M$ occur at lower \neue\  masses and the limit
set by \selR\selR\, searches therefore becomes less stringent. For
$\tan\beta$ values smaller than 2.6, the single electron search takes over,
as long as $\vert \mu \vert$ remains greater than 70\,GeV/$c^2$. The
smallest non-excluded selectron mass, $\mselR = 73$\,GeV/$c^2$, is found in
this region ($\tan\beta=1.5$ and $\mu=-5$\,TeV/$c^2$), for $m_{1/2} =
179$\,GeV/$c^2$ and $m_0 = 2$\,GeV/$c^2$, as displayed in Fig.~\ref{f:m0m2}.

\begin{figure}[h]
\setlength{\unitlength}{1.0cm}
\begin{center}
\begin{picture}(15.0,20.0)
\put(0.0,0.0){\epsfig{file=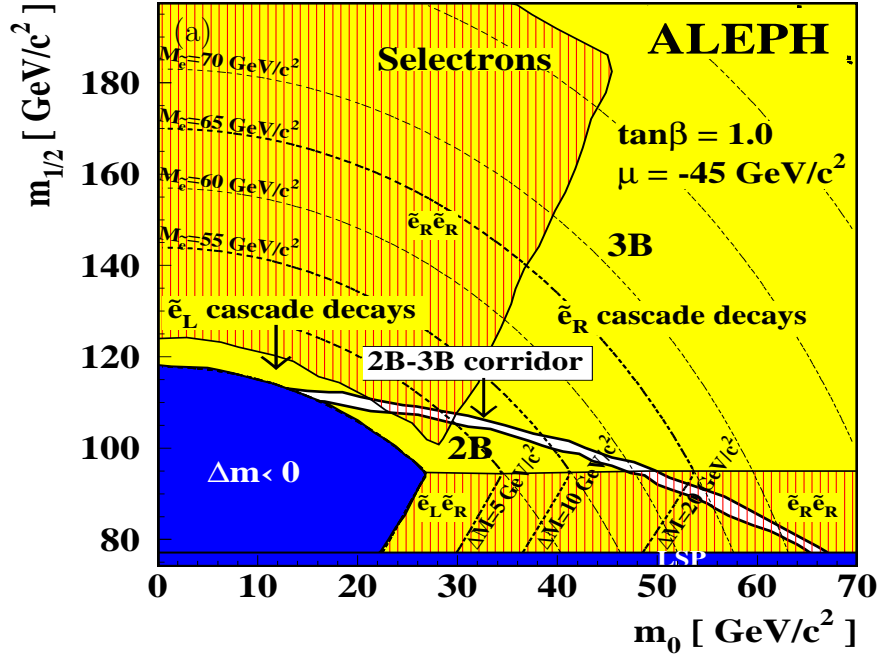,%
            height=10cm,width=12cm}}
\put(0.0,10.0){\epsfig{file=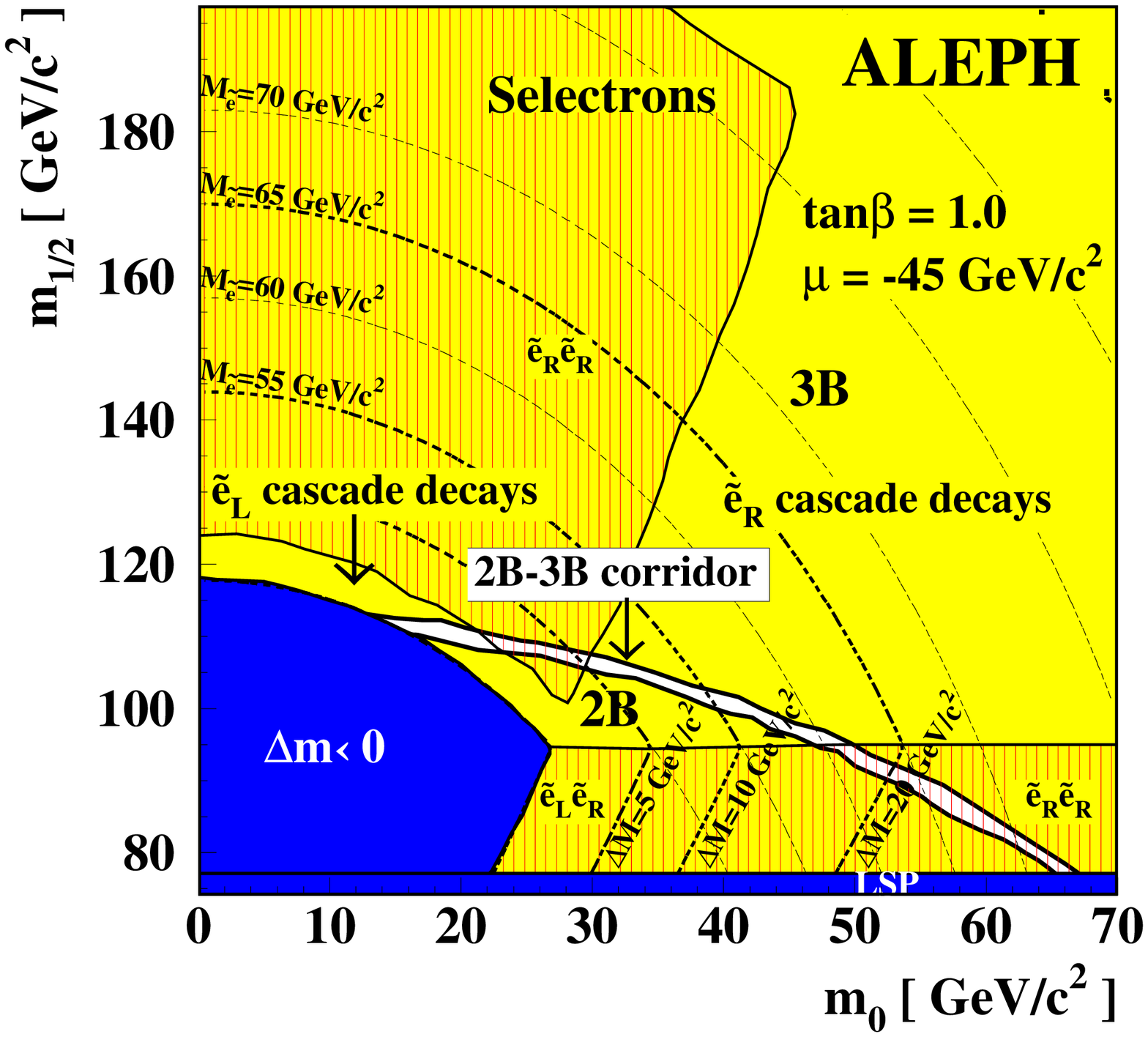,%
            height=10cm,width=12cm}}
\put(2.3,18.5){(a)}
\put(2.3,8.5){(b)}
\end{picture}
\caption{\footnotesize
(a) Regions excluded in the (\mn, \mez) plane by  direct searches for selectrons and charginos, 
 for $\tanb = 1.0$ and $\mu =-45 \mgev$.
The selectron limit at large \mn\ is a line of constant \mez, where the 
photino field content of the \neuz\  becomes large enough 
for the selectrons to decay predominantly via cascades.
The thin dashed curves are the lines of constant $\tilde{\rm e}_{\rm R}$ mass, and the 
thick dashed curves are the lines of constant mass difference with the LSP. Because 
the LSP mass does not depend on \mez\ above 95\,GeV/$c^2$, the iso-\deltm\
lines become identical to the iso-\mselR\ lines.
(b) Same as (a), including the regions that can be covered by dedicated \neue\neud\, 
searches (dark shaded).
The same hatching and shading conventions  as in Fig.~\ref{f:m0m2} are adopted.}
\label{f:gaps}
 \end{center}
 \end{figure}

Small negative values of \Mu\ are in general excluded
by either  chargino searches or  neutralino searches for any selectron mass value.
An example is shown in Fig.~\ref{f:gaps}
where the exclusion domains in the (\mn, \mez) plane are given
for $\tanb = 1.0$ and $\Mu = -45 \mgev$ with and without the dedicated
neutralino searches. It can be seen that the additional \neue\neud\,
searches allow the chargino corridor to be covered.
The overall limit on \mselR\, is therefore set by the
\selR\selL\, analysis at larger negative values of \Mu\, where the \mez\,
coverage of the chargino searches is reduced due to larger chargino and
neutralino masses.

\begin{figure}[h]
\setlength{\unitlength}{1.0cm}
 \begin{center}
\begin{picture}(13.0,13.0)
\put(0.0,0.0){\epsfig{file=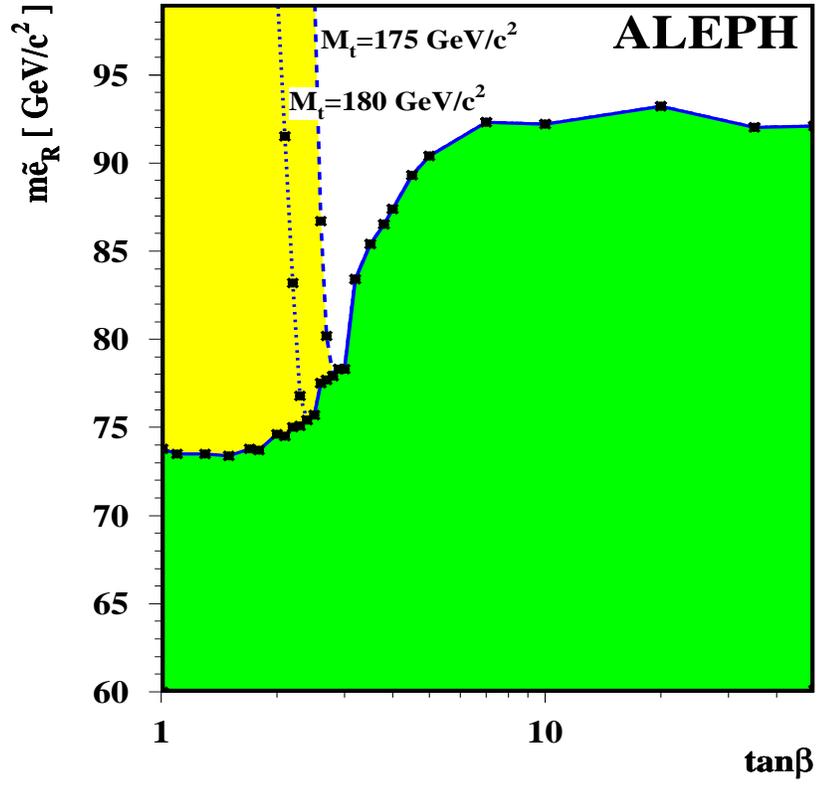,%
            height=13cm,width=13cm}}
\end{picture}
 \caption{ \label{f:limitR}\footnotesize
   Limit on \mselR\,  as a function of $\tanb$. 
  The limit from the Higgs boson searches for  \mtop$=$175(180)\mgev\
 is given by the dashed(dotted) line.}
 \end{center}
\end{figure}

The limit on \mselR\, is shown in Fig.~\ref{f:limitR}
as a function of $\tanb$. Each point represents the result of a
scan
over \Mu, \mez\ and \mn, allowing an absolute lower limit on the selectron
mass to be set at $ \mselR > 73 \mgev.$


\subsection{Limits on \msnu\, and \mselL \label{lresults}}

The present analysis can also be used to derive limits on the masses of the 
heavier selectron and the sneutrino, exploiting the relations given in 
Eqs.~(\ref{eq:selectronL})~and~(\ref{eq:sneutrino}).
For a given value of \tanb,  \mselL\, and \msnu\, take
their minimal values for the same parameter combination, 
because the difference between the 
masses of these two sparticles is only a function of \tanb.
Because the  \mez\, dependence is stronger for \mselL\, and \msnu\, than for \mselR,
the limits are found at smaller values of \mez\, and larger values of \mn. In general,
these limits are located at the intersection of the exclusion
borders of the selectron and  chargino searches or, 
for $\tanb < 1.5$, at the selectron exclusion-border
for \mneue~=~37\mgev\, \cite{char1}.

\begin{figure}[h]
\setlength{\unitlength}{1.0cm}
\begin{center}
\begin{picture}(15.0,20.0)
\put(0.0,0.0){\epsfig{file=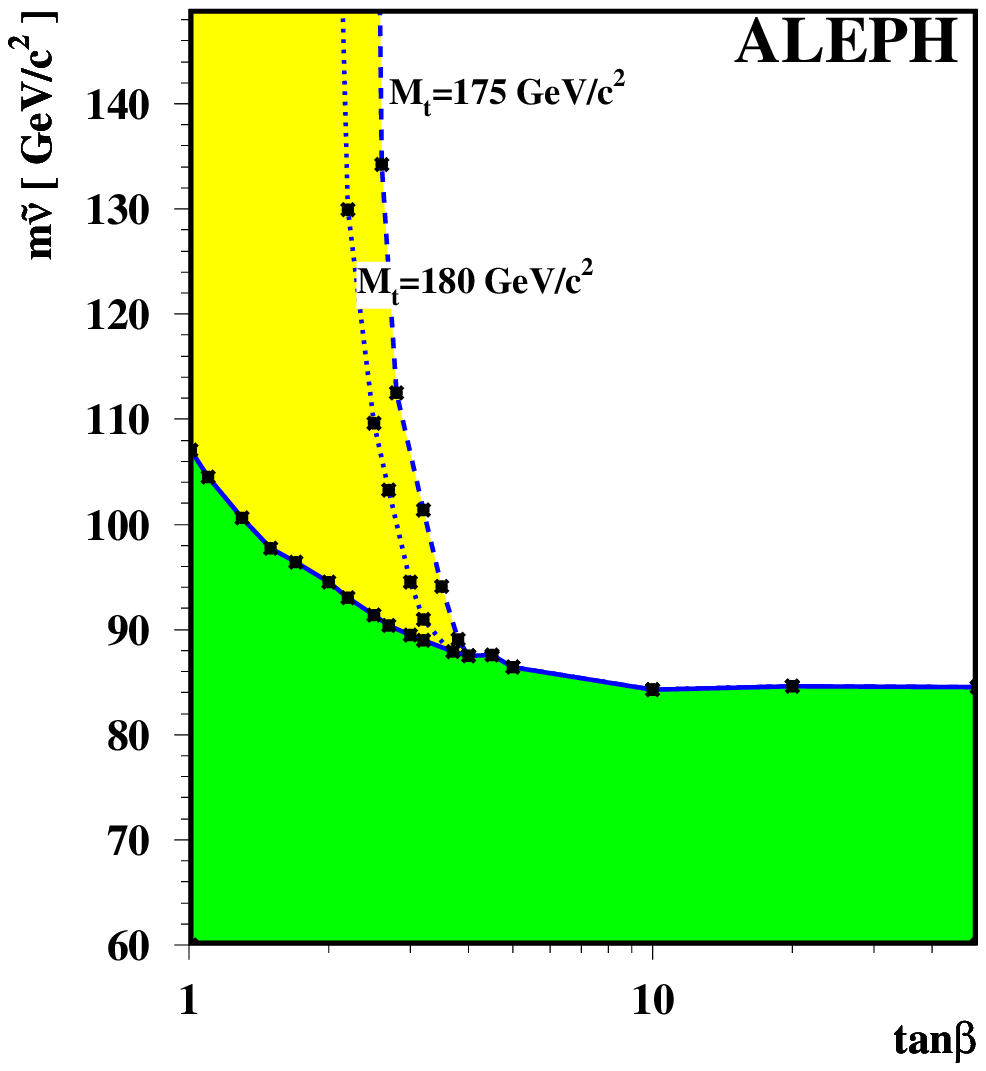,%
            height=11cm,width=12cm}}
\put(0.0,10.0){\epsfig{file=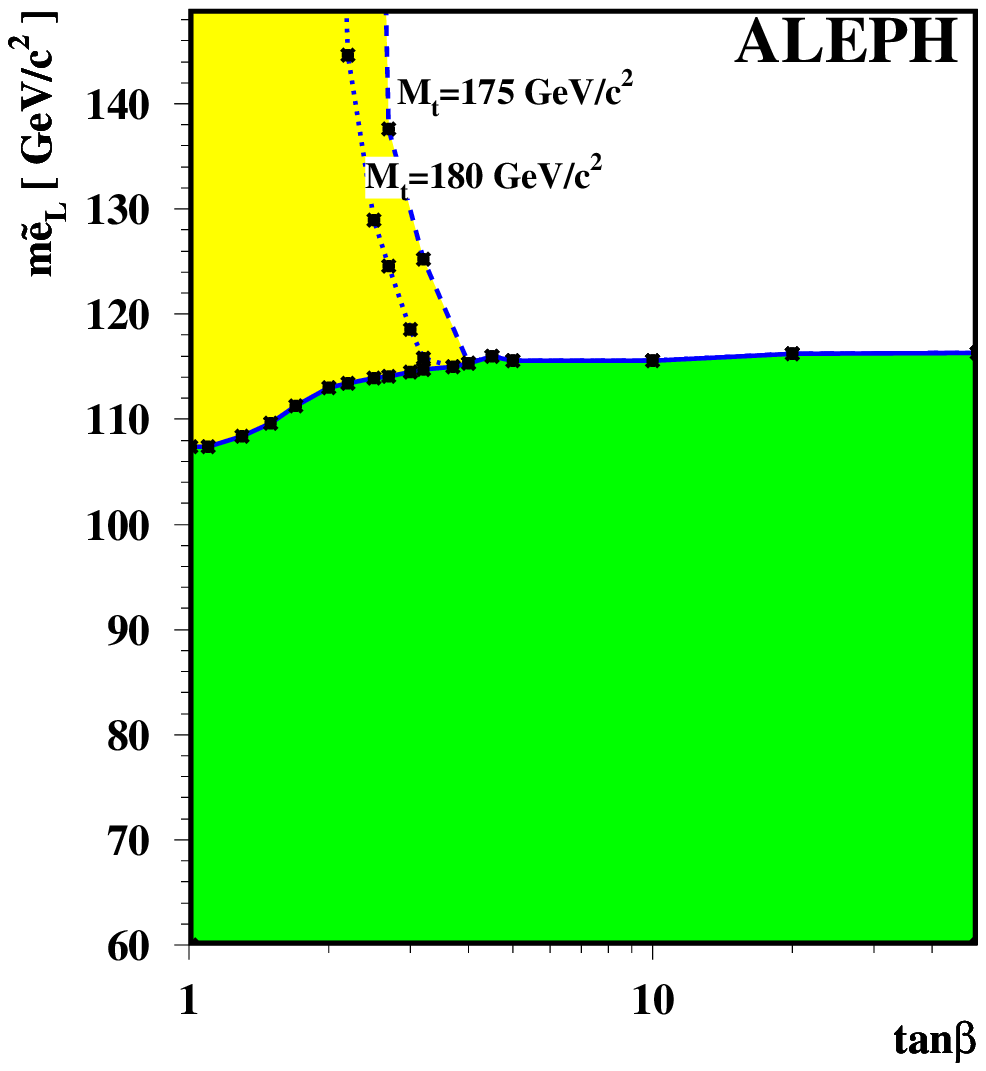,%
            height=11cm,width=12cm}}
\put(2.5,18.7){(a)}
\put(2.5,8.7){(b)}
\end{picture}
 \caption{ \label{f:limitL}\footnotesize
  Limits on \mselL\, (a) and \msnu\, (b)
  as a function of $\tanb$. 
  The limit from the Higgs boson searches for  $\mtop =175(180) \mgev$
 is given by the dashed(dotted) line.}
\end{center} \end{figure}

The lower limits on \mselL\, and \msnu\, are shown in Fig.~\ref{f:limitL} as a function
of \tanb. 
The basic shape reflects the opposite \tanb-dependence of \mselL\, and \msnu. 
The overall limit for the heavier selectron is found to be
$\mselL > 107\mgev$ for $\tanb=1.0$ and $\mu=-80\mgev$. 
For the sneutrinos, an  overall limit of
$\msnu > 84 \mgev $ is obtained   for $\tanb>10$ and $\mu\le-1000\mgev$.


\typeout{*** Higgs ***}
\subsection{Constraints from Higgs boson searches\label{s:higgs}}
As explained  in Ref.~\cite{char1}, 
a lower limit on the mass of the lightest CP-even Higgs boson (\mh)
can be translated into a lower limit on \mez\, as a function of \tanb\, for 
a given value of \mn.
The A boson mass \ma\  and the stop mixing, controlled by 
($\rm A_{t} - \mu\cot\beta$), are chosen in a way that maximizes \mh\, for a
given set of
\mez, \mn\, and \tanb.
The limit on \mez\, decreases with increasing
\mn.  Therefore, the choice \mn~=~100~\mgev, corresponding to the kinematic
limit for selectron production 
(Eq.~\ref{eq:selectronR}), is conservative for the 
present analysis. 

The  limit becomes less stringent with 
increasing \mtop. The impact of the \mtop\, uncertainty ($\pm 5\mgev$)
is estimated by performing the calculation for $\mtop = 175$ and $180\mgev$.
The results are obtained with the ALEPH lower limit on \mh~\cite{higgs}.
The limit on \mez\, decreases rapidly with increasing \tanb, hence the impact of the
Higgs boson searches on the present 
results is sizeable only for small \tanb. 

The resulting lower limits on \mselR, \mselL\, and \msnu\, are shown in 
Figs.~\ref{f:limitR}~and~\ref{f:limitL} for \mtop~=~175~(180)\mgev.
The overall limit on \mselR\, is found at $\tanb=2.8(2.4)$,
$ \mselR > 77(75)\mgev.$
The limit on \mselL\, is found to be
$\mselL > 115(115)\mgev$.
Since the limit on \msnu\, is found at large \tanb,
the constraints from Higgs boson searches have no impact in this case.


\subsection{Interpretation in mSUGRA \label{s:msugra}}

The results of the searches for selectrons, charginos, neutralinos and Higgs
bosons are also combined within the framework of minimal supergravity, 
following the analysis presented in Ref.~\cite{char1}. 
Scans of the $(m_0, m_{1/2})$ plane are performed as a function of $\tanb$, 
for both signs of $\mu$ and for $A_0 = 0$. 

The results of the neutral Higgs boson searches [5] are interpreted for
hZ, HZ and hA production, where h and H are, respectively, the lighter and the
heavier CP-even Higgs bosons, and A the CP-odd Higgs boson. 

\begin{figure}[h]
\setlength{\unitlength}{1.0cm}
\begin{center}
\begin{picture}(16.0,19.0)
\put(-0.3,6.5){\epsfig{file=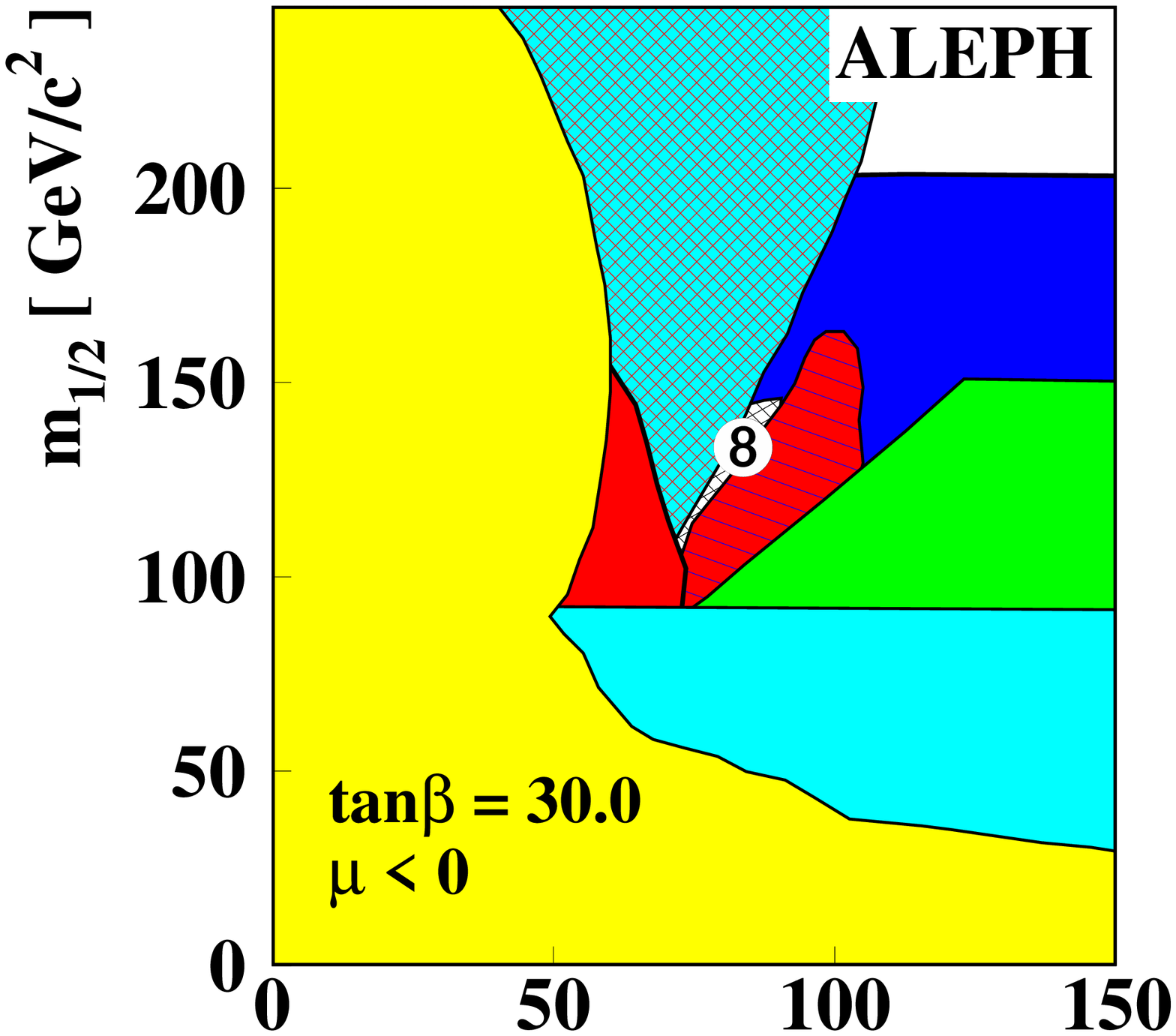,%
            height=7.5cm,width=9.0cm}}
\put(-0.3,13.0){\epsfig{file=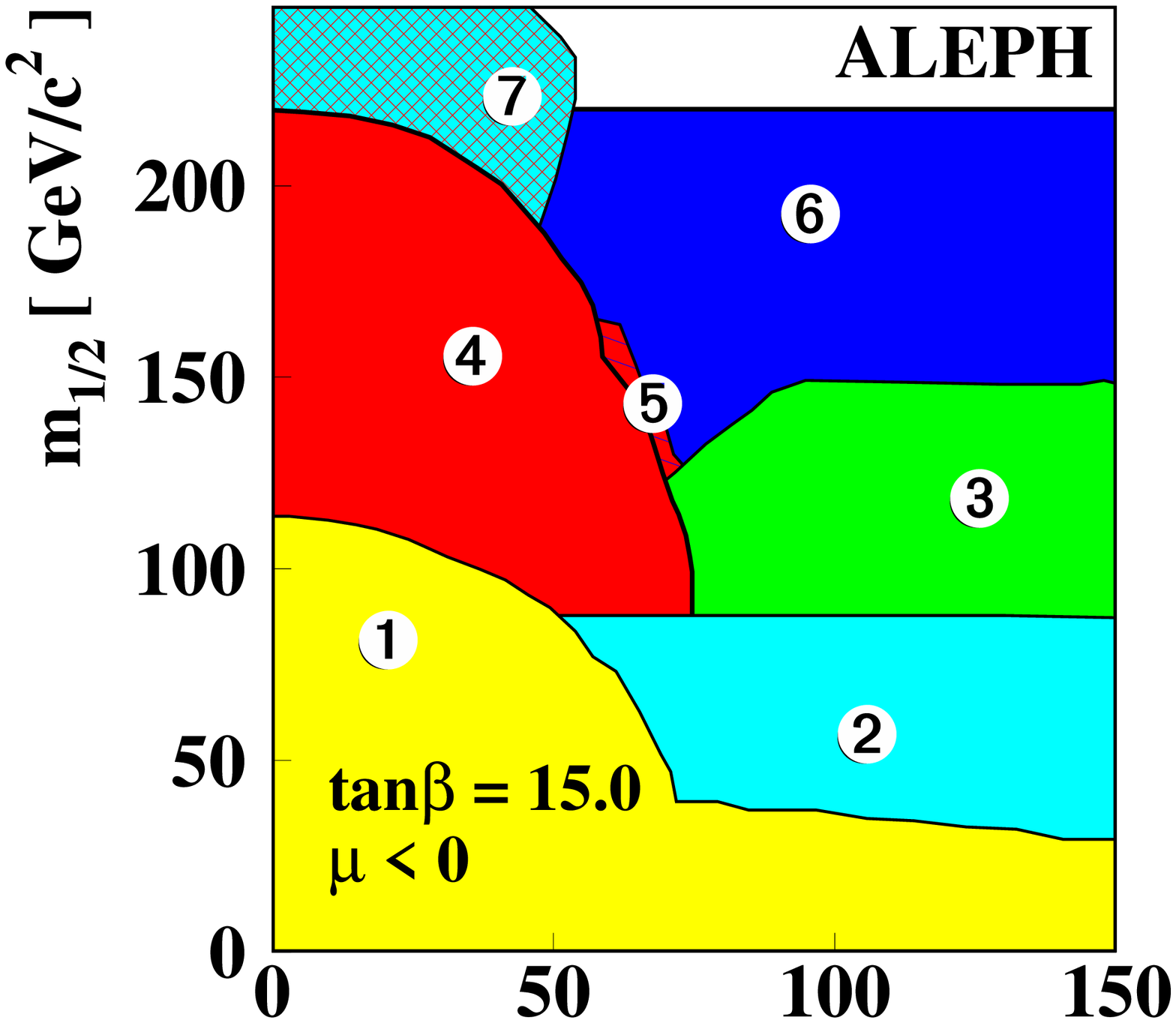,%
            height=7.5cm,width=9.0cm}}
\put(8.2,6.5){\epsfig{file=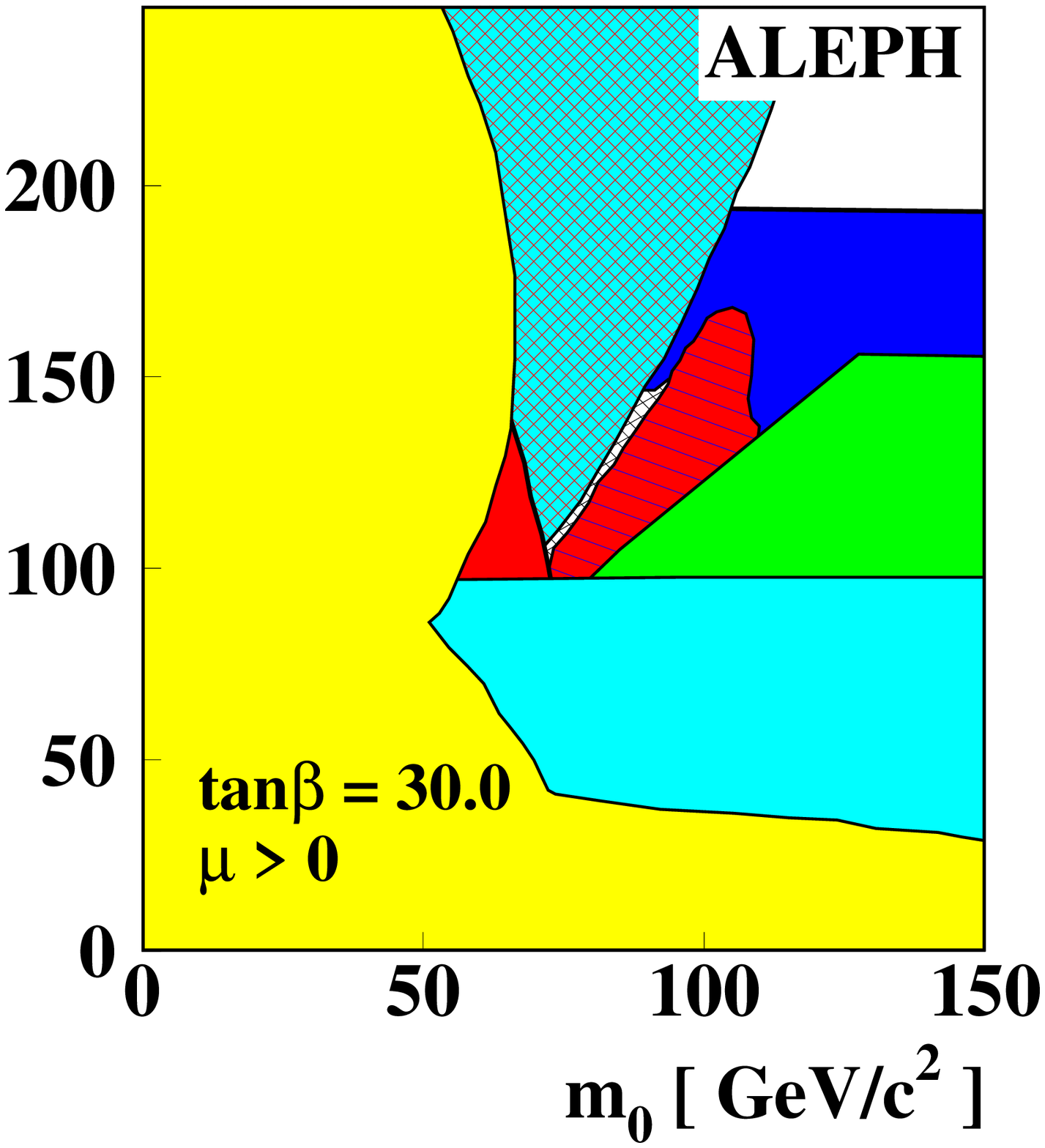,%
            height=7.5cm,width=9.0cm}}
\put(8.2,13.0){\epsfig{file=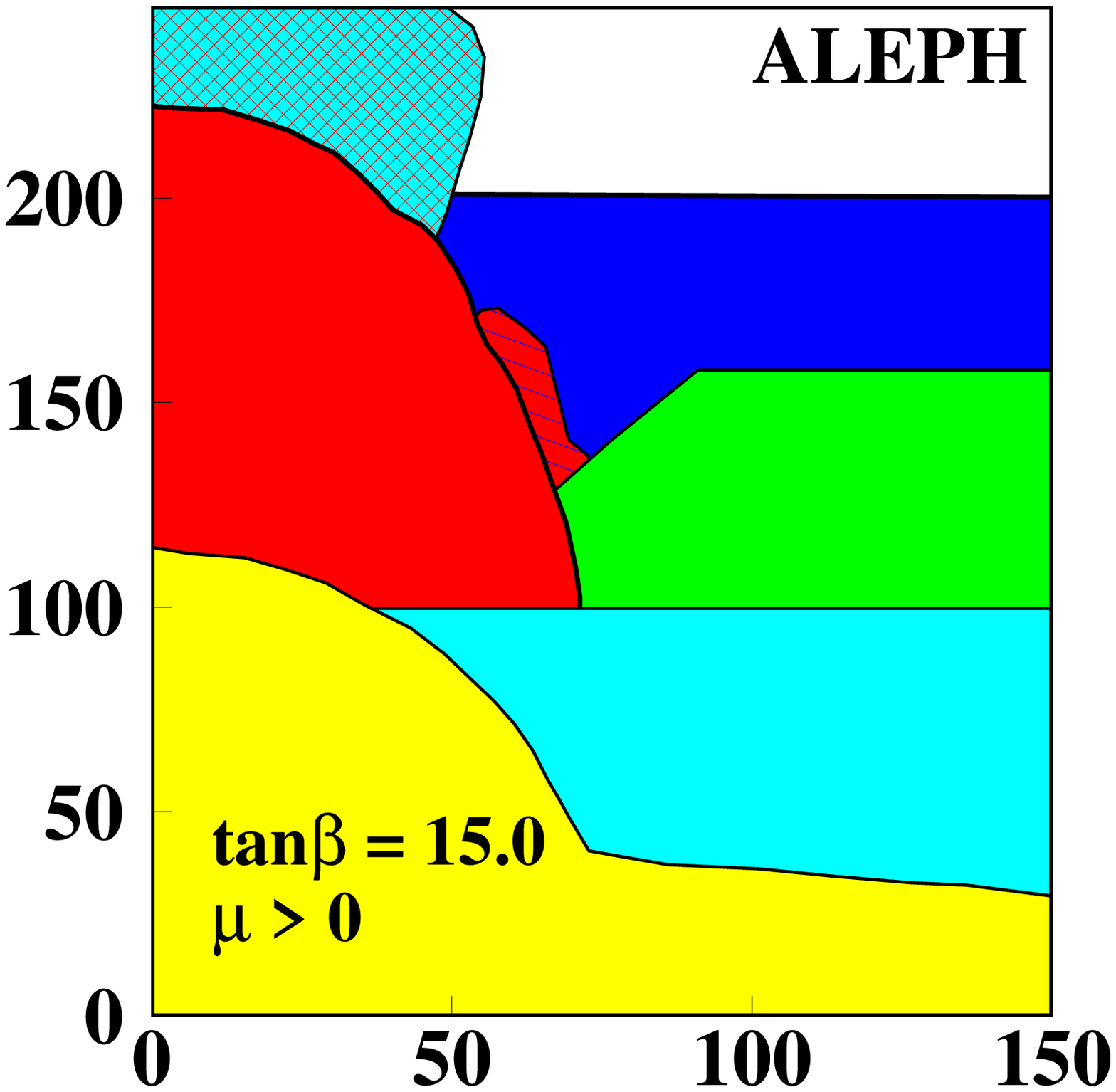,%
            height=7.5cm,width=9.0cm}}
\put(-0.3,0.0){\epsfig{file=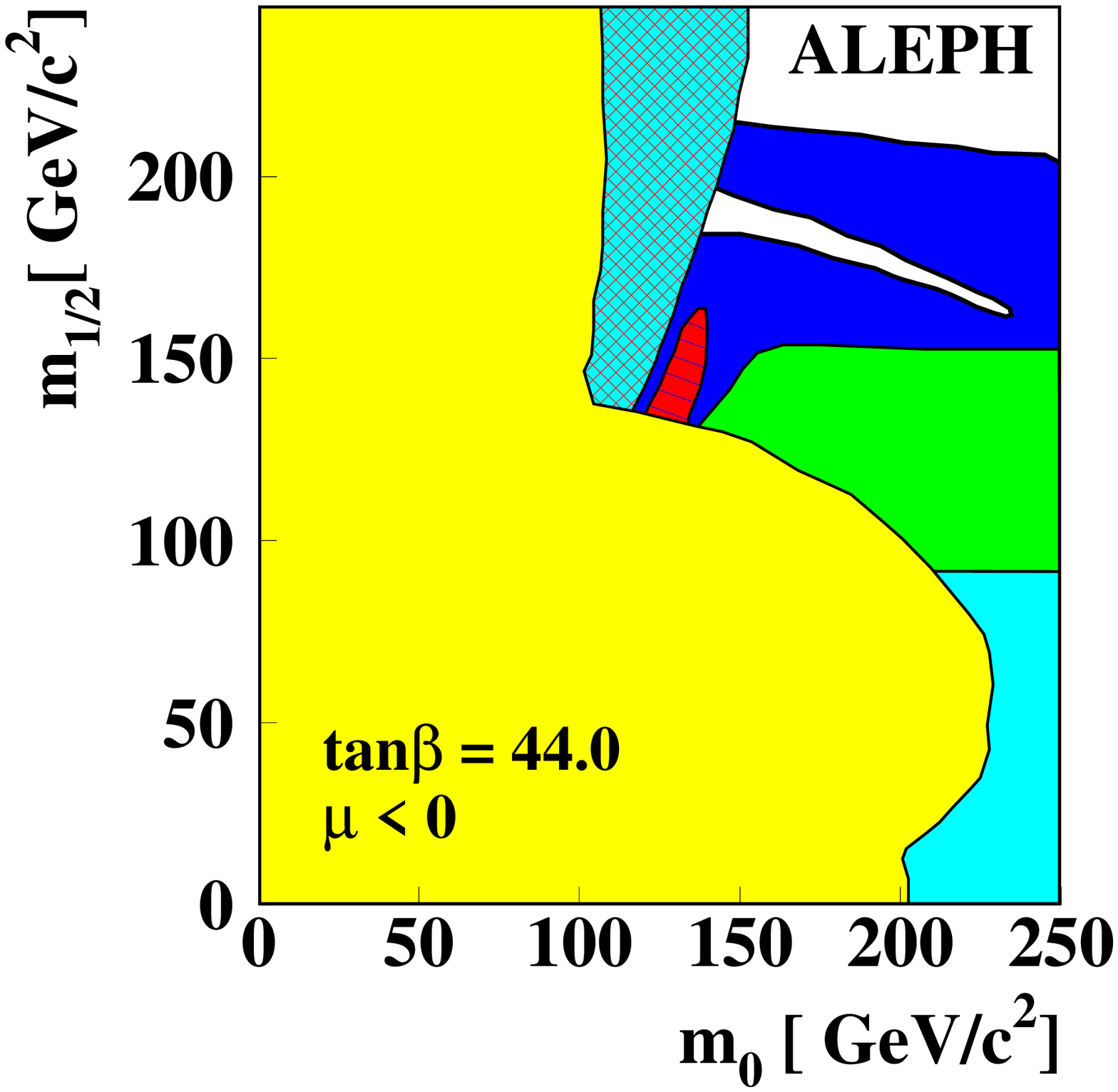,%
            height=7.5cm,width=9.0cm}}
\put(1.8,19.35){(a)}
\put(10.3,19.35){(b)}
\put(1.8,12.8){(c)}
\put(10.3,12.8){(d)}
\put(1.8,6.3){(e)}
\end{picture}
 \caption{ \label{f:sugra_m0m2}\footnotesize
Minimal Supergravity scenario: 
regions excluded in the (\mn, \mez) plane 
for $\tanb = 15$, 30 and 44 and for $\An = 0$.
Region 1 is theoretically forbidden.
The other regions are excluded by LEP1 (2), and by searches for 
charginos (3), selectrons (4), staus (5), Higgs bosons (6), 
stable charged particles (7), and associated neutralino production (8).}
\end{center} \end{figure}

In minimal supergravity, the trilinear couplings $A_\tau$, $A_{\rm b}$ and 
$A_{\rm t}$ are unambiguously predicted from the model parameters. Mixing 
effects can therefore not be arbitrarily switched off as is done in the 
previous sections. While mixing in the squark sector is relevant only 
for the Higgs boson mass and coupling prediction, mixing in the stau sector 
can lead to a stau much lighter than selectrons and smuons, thus affecting
the decay phenomenology of charginos, neutralinos and Higgs bosons. Decays
into staus may indeed become predominant and lead to final state topologies 
with taus. These topologies are not always efficiently covered by the searches 
described in the previous sections, in particular if the mass difference between 
the lightest neutralino and the stau is small. Three additional 
searches are used to address this new situation:
\begin{enumerate}
\item the search for the associated neutralino production ${\rm e}^+{\rm e}^-
\to \chi_2^0 \chi_1^0$ with the subsequent decay $\chi_2^0 \to
\tilde\tau \tau \to \chi_1^0 \tau \tau$~\cite{char2},
leading to at least one visible $\tau$ in the final state;
\item the search for an invisible Higgs boson~\cite{higgs}, which covers the
${\rm e}^+{\rm e}^- \to {\rm hZ}$ process followed by the decay
$\rm h \to \tilde\tau\tilde\tau \to \chi_1^0 \chi_1^0 \tau \tau$;
\item the search for heavy stable charged particles~\cite{gmsb} which addresses
the stau-pair production ${\rm e}^+{\rm e}^- \to \tilde\tau\tilde\tau$
when the mass difference with the LSP is smaller than $m_\tau$.
\end{enumerate}
The impact of each of the analyses (standard and additional) in the $(m_0, m_{1/2})$ 
plane is illustrated in Figs.~\ref{f:sugra_m0m2}a to~\ref{f:sugra_m0m2}d for two 
typical $\tan\beta$ values, $\tan\beta =$ 15 and 30, and for both signs of $\mu$.
In general, small \mez\ values are excluded by selectron and Higgs boson searches. 
Small \mn\ values either are theoretically forbidden or correspond to a stau LSP. 

\begin{figure}[h]
\setlength{\unitlength}{1.0cm}
\begin{center}
\begin{picture}(16.0,17.0)
\put(-0.5,0.5){\epsfig{file=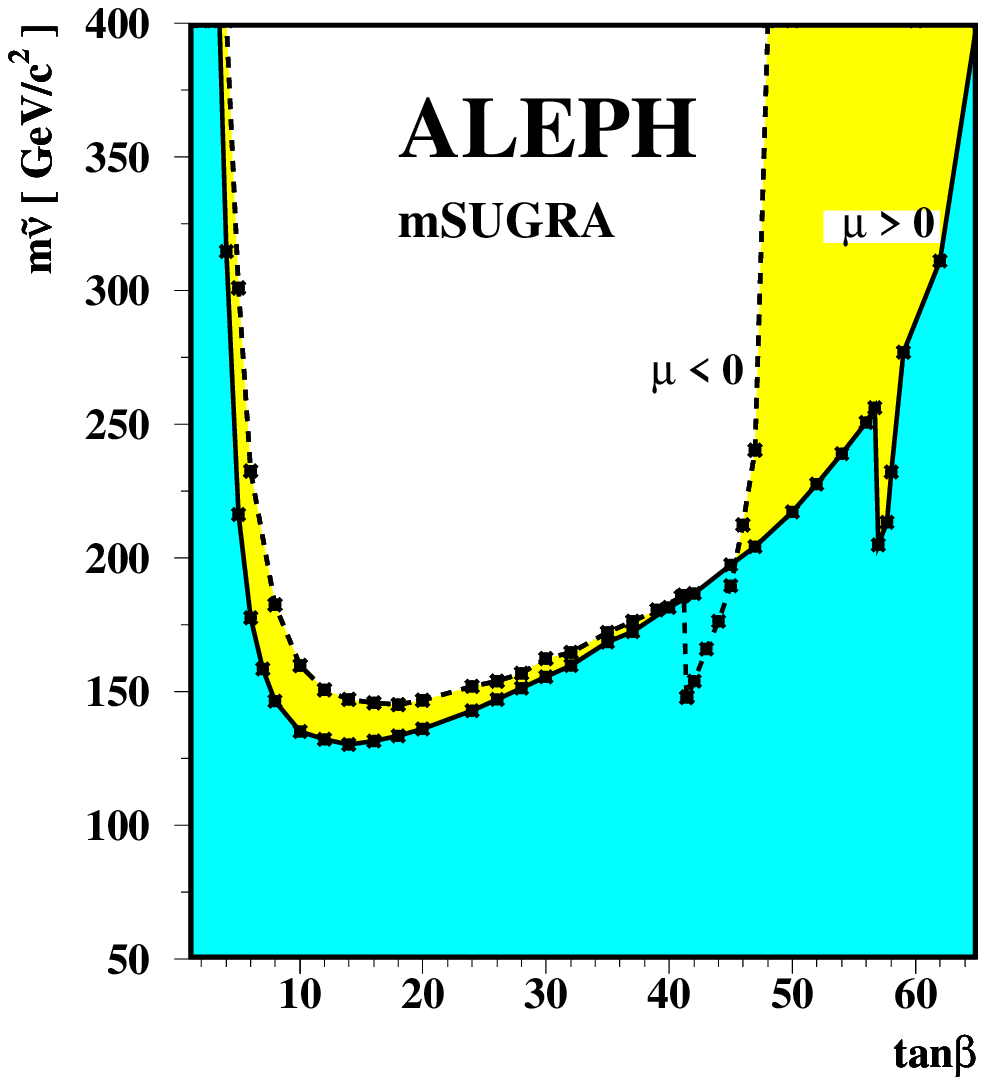,%
            height=10.0cm,width=10.0cm}}
\put(-0.5,9.0){\epsfig{file=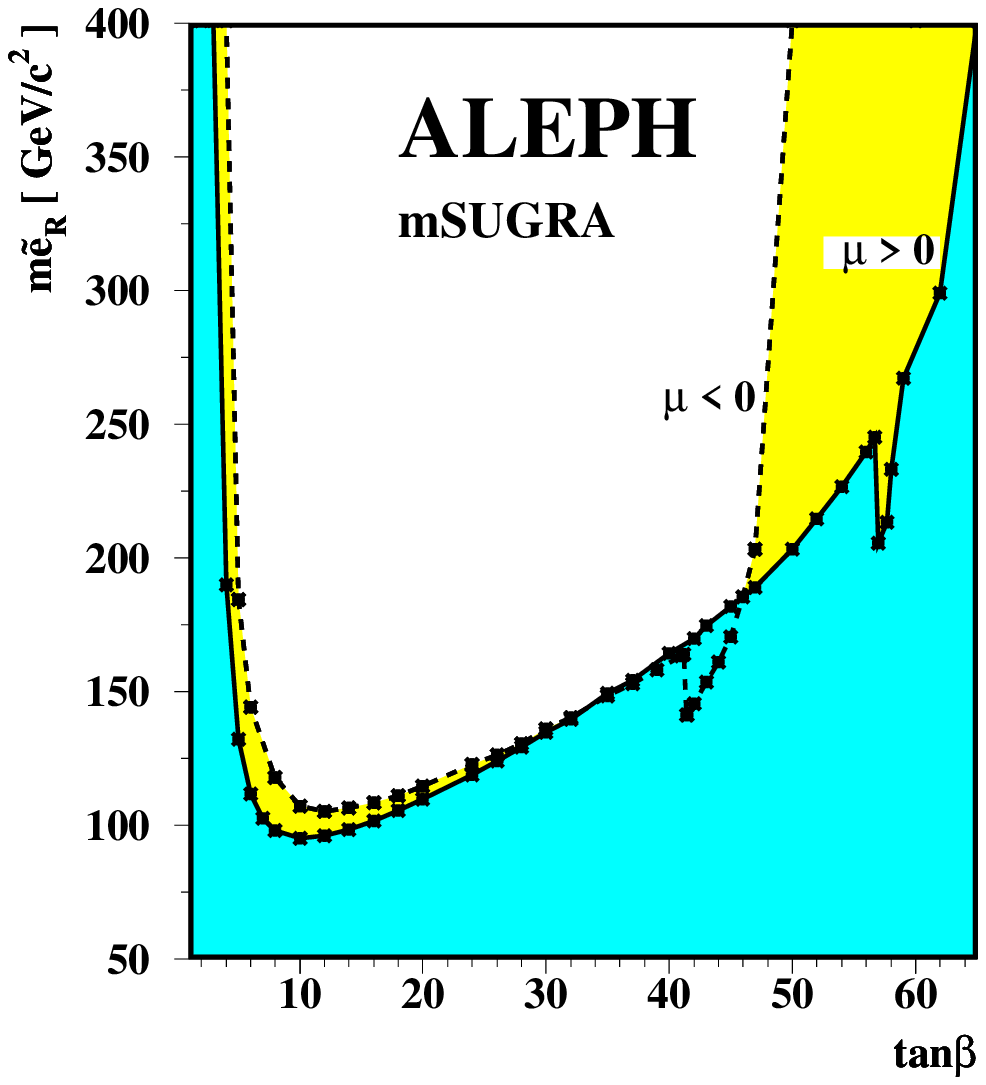,%
            height=10.0cm,width=10.0cm}}
\put(8.0,9.0){\epsfig{file=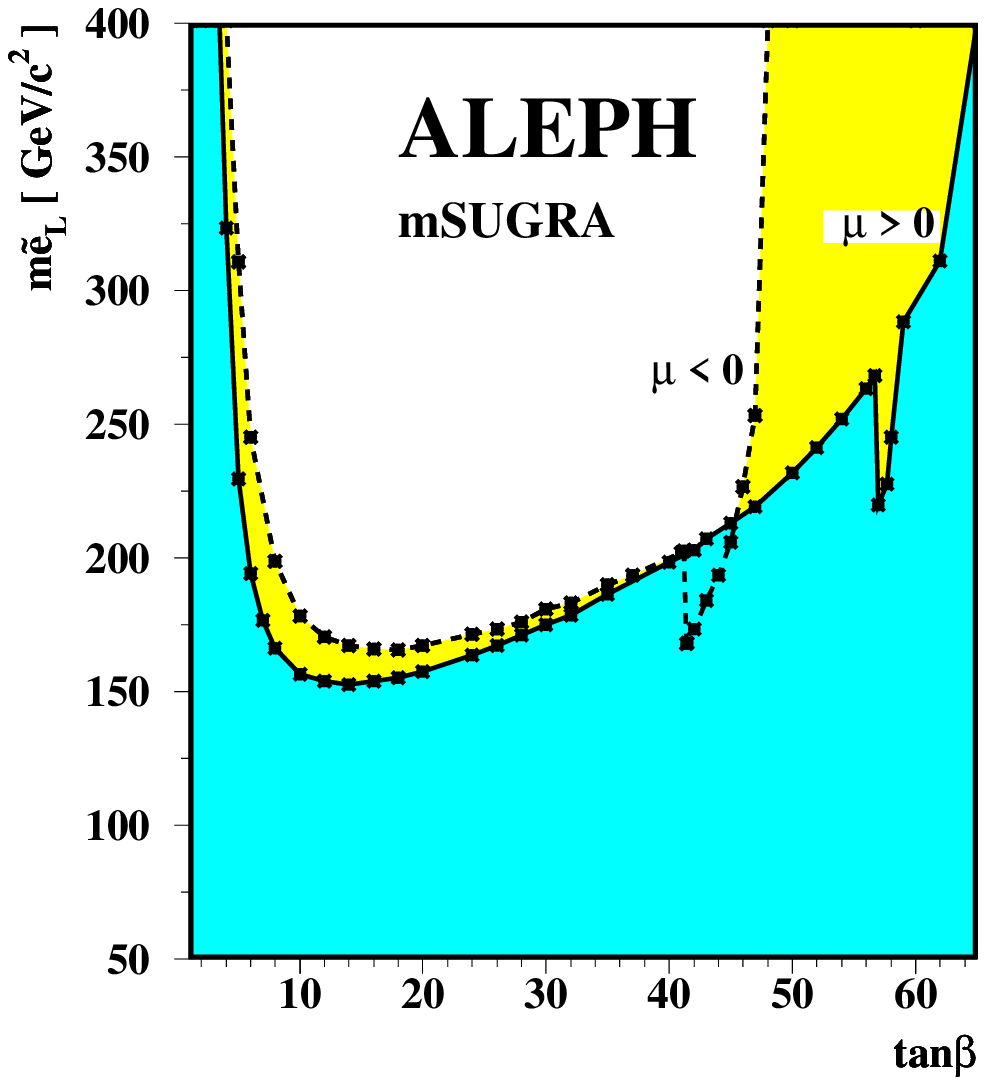,%
            height=10.0cm,width=10.0cm}}
\put(1.7,17.0){(a)}
\put(10.3,17.0){(b)}
\put(1.7,8.5){(c)}
\end{picture}
 \caption{ \label{f:lim_sugra}\footnotesize
   Limits within mSUGRA on \mselR\, (a), \mselL\, (b)  
   and \msnu\, (c) as a function of $\tanb$ 
   for $\Mu<0$ (dashed line) and $\Mu>0$
   (full line). Detailed explanations about 
the structure at large \tanb\ values can be found in 
Section~\ref{s:msugra}.
} 

\end{center} \end{figure}

The excluded domains in the $(m_0, m_{1/2})$ plane can be translated into 
lower limits on \mselR, \mselL\ and $\msnu$, shown in Fig.~\ref{f:lim_sugra} 
as a function of \tanb, for $\mtop = 175$\,GeV/$c^2$ and for both signs of $\mu$. 
With increasing \tanb, the domains covered by Higgs boson searches shrink, while 
the regions with a stau LSP extend because of mixing effects in the stau sector.
The minimal combined exclusion is reached for intermediate \tanb\ values, around~15.
The structure at large \tanb\ is due to a loss of sensitivity of the combined search 
for the hZ and HZ processes. As can be seen in the example shown in 
Fig.~\ref{f:sugra_m0m2}e, a non-excluded channel of the 
$(m_0, m_{1/2})$ plane opens up, in which the hZ coupling is too small and the H mass too
large for the hZ and HZ production to contribute significantly. 

Altogether, a lower limit on the selectron mass of 95\,\mgev\ is derived for
$A_0 = 0$ and for both signs of $\mu$. For the heavier selectron and the sneutrino, mass lower
limits of 152 and 130\,\mgev\ are obtained, respectively. 


\section{Conclusions}

The results of searches for selectron,
chargino and neutralino production in the data collected 
by ALEPH at centre-of-mass energies up to 209~GeV have been
interpreted in the framework of the MSSM with R-parity conservation,
gaugino and sfermion mass unification, and no sfermion mixing.
A scan over the four parameters
$\tanb, \Mu, \mez$ and \mn\, has been performed 
to determine the following lower limits on the masses of selectrons and  sneutrinos:
\[\mselR > 73\mgev,\] 
\[\mselL > 107\mgev,\] 
\[\msnu > 84 \mgev.\]
These limits improve on earlier 
results obtained at lower centre-of-mass energies by the L3 collaboration \cite{l3lep}.\\

{\noindent The limits on the selectron masses can be further improved by 
including constraints from Higgs boson searches. The results 
depend slightly on the value of the top quark mass.
For \mtop~=~175(180)~\mgev, mass limits of
\[\mselR > 77(75)\mgev,\]
\[\mselL > 115(115)\mgev\]
are obtained.}\\

{\noindent Within minimal supergravity, mass lower limits have been set at
\[\mselR > 95\mgev,\]
\[\mselL > 152\mgev,\]
\[\msnu > 130 \mgev,\]
for $\An = 0$ and $\mtop = 175$\,GeV/$c^2$.}

\section*{Acknowledgements}

It is a pleasure to congratulate our colleagues from the
accelerator divisions for the outstanding operation of LEP\,2, especially in
its last year of running during which the accelerator performance were
pushed beyond expectation. We are indebted to the engineers and technicians
in all our institutions for their contributions to the excellent performance
of ALEPH. Those of us from non-member states wish to thank CERN for its
hospitality and support.





\end{document}

%% file: authb.tex
\pagestyle{empty}
\newpage
\small
%
%
\newlength{\saveparskip}
\newlength{\savetextheight}
\newlength{\savetopmargin}
\newlength{\savetextwidth}
\newlength{\saveoddsidemargin}
\newlength{\savetopsep}
\setlength{\saveparskip}{\parskip}
\setlength{\savetextheight}{\textheight}
\setlength{\savetopmargin}{\topmargin}
\setlength{\savetextwidth}{\textwidth}
\setlength{\saveoddsidemargin}{\oddsidemargin}
\setlength{\savetopsep}{\topsep}
%
%
\setlength{\parskip}{0.0cm}
\setlength{\textheight}{25.0cm}
\setlength{\topmargin}{-1.5cm}
\setlength{\textwidth}{16 cm}
\setlength{\oddsidemargin}{-0.0cm}
\setlength{\topsep}{1mm}
\pretolerance=10000
\centerline{\large\bf The ALEPH Collaboration}
\footnotesize
\vspace{0.5cm}
{\raggedbottom
\begin{sloppypar}
\samepage\noindent
A.~Heister,
S.~Schael
\nopagebreak
\begin{center}
\parbox{15.5cm}{\sl\samepage
Physikalisches Institut das RWTH-Aachen, D-52056 Aachen, Germany}
\end{center}\end{sloppypar}
\vspace{2mm}
\begin{sloppypar}
\noindent
R.~Barate,
R.~Bruneli\`ere,
I.~De~Bonis,
D.~Decamp,
C.~Goy,
S.~Jezequel,
J.-P.~Lees,
F.~Martin,
E.~Merle,
\mbox{M.-N.~Minard},
B.~Pietrzyk,
B.~Trocm\'e
\nopagebreak
\begin{center}
\parbox{15.5cm}{\sl\samepage
Laboratoire de Physique des Particules (LAPP), IN$^{2}$P$^{3}$-CNRS,
F-74019 Annecy-le-Vieux Cedex, France}
\end{center}\end{sloppypar}
\vspace{2mm}
\begin{sloppypar}
\noindent
G.~Boix,$^{25}$
S.~Bravo,
M.P.~Casado,
M.~Chmeissani,
J.M.~Crespo,
E.~Fernandez,
M.~Fernandez-Bosman,
Ll.~Garrido,$^{15}$
E.~Graug\'{e}s,
J.~Lopez,
M.~Martinez,
G.~Merino,
A.~Pacheco,
D.~Paneque,
H.~Ruiz
\nopagebreak
\begin{center}
\parbox{15.5cm}{\sl\samepage
Institut de F\'{i}sica d'Altes Energies, Universitat Aut\`{o}noma
de Barcelona, E-08193 Bellaterra (Barcelona), Spain$^{7}$}
\end{center}\end{sloppypar}
\vspace{2mm}
\begin{sloppypar}
\noindent
A.~Colaleo,
D.~Creanza,
N.~De~Filippis,
M.~de~Palma,
G.~Iaselli,
G.~Maggi,
M.~Maggi,
S.~Nuzzo,
A.~Ranieri,
G.~Raso,$^{24}$
F.~Ruggieri,
G.~Selvaggi,
L.~Silvestris,
P.~Tempesta,
A.~Tricomi,$^{3}$
G.~Zito
\nopagebreak
\begin{center}
\parbox{15.5cm}{\sl\samepage
Dipartimento di Fisica, INFN Sezione di Bari, I-70126 Bari, Italy}
\end{center}\end{sloppypar}
\vspace{2mm}
\begin{sloppypar}
\noindent
X.~Huang,
J.~Lin,
Q. Ouyang,
T.~Wang,
Y.~Xie,
R.~Xu,
S.~Xue,
J.~Zhang,
L.~Zhang,
W.~Zhao
\nopagebreak
\begin{center}
\parbox{15.5cm}{\sl\samepage
Institute of High Energy Physics, Academia Sinica, Beijing, The People's
Republic of China$^{8}$}
\end{center}\end{sloppypar}
\vspace{2mm}
\begin{sloppypar}
\noindent
D.~Abbaneo,
P.~Azzurri,
T.~Barklow,$^{30}$
O.~Buchm\"uller,$^{30}$
M.~Cattaneo,
F.~Cerutti,
B.~Clerbaux,$^{23}$
H.~Drevermann,
R.W.~Forty,
M.~Frank,
F.~Gianotti,
T.C.~Greening,$^{26}$
J.B.~Hansen,
J.~Harvey,
D.E.~Hutchcroft,
P.~Janot,
B.~Jost,
M.~Kado,$^{2}$
P.~Mato,
A.~Moutoussi,
F.~Ranjard,
L.~Rolandi,
D.~Schlatter,
G.~Sguazzoni,
W.~Tejessy,
F.~Teubert,
A.~Valassi,
I.~Videau,
J.J.~Ward
\nopagebreak
\begin{center}
\parbox{15.5cm}{\sl\samepage
European Laboratory for Particle Physics (CERN), CH-1211 Geneva 23,
Switzerland}
\end{center}\end{sloppypar}
\vspace{2mm}
\begin{sloppypar}
\noindent
F.~Badaud,
S.~Dessagne,
A.~Falvard,$^{20}$
D.~Fayolle,
P.~Gay,
J.~Jousset,
B.~Michel,
S.~Monteil,
D.~Pallin,
J.M.~Pascolo,
P.~Perret
\nopagebreak
\begin{center}
\parbox{15.5cm}{\sl\samepage
Laboratoire de Physique Corpusculaire, Universit\'e Blaise Pascal,
IN$^{2}$P$^{3}$-CNRS, Clermont-Ferrand, F-63177 Aubi\`{e}re, France}
\end{center}\end{sloppypar}
\vspace{2mm}
\begin{sloppypar}
\noindent
J.D.~Hansen,
J.R.~Hansen,
P.H.~Hansen,
B.S.~Nilsson
\nopagebreak
\begin{center}
\parbox{15.5cm}{\sl\samepage
Niels Bohr Institute, 2100 Copenhagen, DK-Denmark$^{9}$}
\end{center}\end{sloppypar}
\vspace{2mm}
\begin{sloppypar}
\noindent
A.~Kyriakis,
C.~Markou,
E.~Simopoulou,
A.~Vayaki,
K.~Zachariadou
\nopagebreak
\begin{center}
\parbox{15.5cm}{\sl\samepage
Nuclear Research Center Demokritos (NRCD), GR-15310 Attiki, Greece}
\end{center}\end{sloppypar}
\vspace{2mm}
\begin{sloppypar}
\noindent
A.~Blondel,$^{12}$
\mbox{J.-C.~Brient},
F.~Machefert,
A.~Roug\'{e},
M.~Swynghedauw,
R.~Tanaka
\linebreak
H.~Videau
\nopagebreak
\begin{center}
\parbox{15.5cm}{\sl\samepage
Laoratoire Leprince-Ringuet, Ecole
Polytechnique, IN$^{2}$P$^{3}$-CNRS, \mbox{F-91128} Palaiseau Cedex, France}
\end{center}\end{sloppypar}
\vspace{2mm}
\begin{sloppypar}
\noindent
V.~Ciulli,
E.~Focardi,
G.~Parrini
\nopagebreak
\begin{center}
\parbox{15.5cm}{\sl\samepage
Dipartimento di Fisica, Universit\`a di Firenze, INFN Sezione di Firenze,
I-50125 Firenze, Italy}
\end{center}\end{sloppypar}
\vspace{2mm}
\begin{sloppypar}
\noindent
A.~Antonelli,
M.~Antonelli,
G.~Bencivenni,
F.~Bossi,
G.~Capon,
V.~Chiarella,
P.~Laurelli,
G.~Mannocchi,$^{5}$
G.P.~Murtas,
L.~Passalacqua
\nopagebreak
\begin{center}
\parbox{15.5cm}{\sl\samepage
Laboratori Nazionali dell'INFN (LNF-INFN), I-00044 Frascati, Italy}
\end{center}\end{sloppypar}
\vspace{2mm}
\begin{sloppypar}
\noindent
J.~Kennedy,
J.G.~Lynch,
P.~Negus,
V.~O'Shea,
A.S.~Thompson
\nopagebreak
\begin{center}
\parbox{15.5cm}{\sl\samepage
Department of Physics and Astronomy, University of Glasgow, Glasgow G12
8QQ,United Kingdom$^{10}$}
\end{center}\end{sloppypar}
\vspace{2mm}
\begin{sloppypar}
\noindent
S.~Wasserbaech
\nopagebreak
\begin{center}
\parbox{15.5cm}{\sl\samepage
Department of Physics, Haverford College, Haverford, PA 19041-1392, U.S.A.}
\end{center}\end{sloppypar}
\vspace{2mm}
\begin{sloppypar}
\noindent
R.~Cavanaugh,$^{4}$
S.~Dhamotharan,$^{21}$
C.~Geweniger,
P.~Hanke,
V.~Hepp,
E.E.~Kluge,
G.~Leibenguth,
A.~Putzer,
H.~Stenzel,
K.~Tittel,
M.~Wunsch$^{19}$
\nopagebreak
\begin{center}
\parbox{15.5cm}{\sl\samepage
Kirchhoff-Institut f\"ur Physik, Universit\"at Heidelberg, D-69120
Heidelberg, Germany$^{16}$}
\end{center}\end{sloppypar}
\vspace{2mm}
\begin{sloppypar}
\noindent
R.~Beuselinck,
W.~Cameron,
G.~Davies,
P.J.~Dornan,
M.~Girone,$^{1}$
R.D.~Hill,
N.~Marinelli,
J.~Nowell,
S.A.~Rutherford,
J.K.~Sedgbeer,
J.C.~Thompson,$^{14}$
R.~White
\nopagebreak
\begin{center}
\parbox{15.5cm}{\sl\samepage
Department of Physics, Imperial College, London SW7 2BZ,
United Kingdom$^{10}$}
\end{center}\end{sloppypar}
\vspace{2mm}
\begin{sloppypar}
\noindent
V.M.~Ghete,
P.~Girtler,
E.~Kneringer,
D.~Kuhn,
G.~Rudolph
\nopagebreak
\begin{center}
\parbox{15.5cm}{\sl\samepage
Institut f\"ur Experimentalphysik, Universit\"at Innsbruck, A-6020
Innsbruck, Austria$^{18}$}
\end{center}\end{sloppypar}
\vspace{2mm}
\begin{sloppypar}
\noindent
E.~Bouhova-Thacker,
C.K.~Bowdery,
D.P.~Clarke,
G.~Ellis,
A.J.~Finch,
F.~Foster,
G.~Hughes,
R.W.L.~Jones,
M.R.~Pearson,
N.A.~Robertson,
M.~Smizanska
\nopagebreak
\begin{center}
\parbox{15.5cm}{\sl\samepage
Department of Physics, University of Lancaster, Lancaster LA1 4YB,
United Kingdom$^{10}$}
\end{center}\end{sloppypar}
\vspace{2mm}
\begin{sloppypar}
\noindent
O.~van~der~Aa,
C.~Delaere,
V.~Lemaitre
\nopagebreak
\begin{center}
\parbox{15.5cm}{\sl\samepage
Institut de Physique Nucl\'eaire, D\'epartement de Physique, Universit\'e Catholique de Louvain, 1348 Louvain-la-Neuve, Belgium}
\end{center}\end{sloppypar}
\vspace{2mm}
\begin{sloppypar}
\noindent
U.~Blumenschein,
F.~H\"olldorfer,
K.~Jakobs,
F.~Kayser,
K.~Kleinknecht,
A.-S.~M\"uller,
G.~Quast,$^{6}$
B.~Renk,
H.-G.~Sander,
S.~Schmeling,
H.~Wachsmuth,
C.~Zeitnitz,
T.~Ziegler
\nopagebreak
\begin{center}
\parbox{15.5cm}{\sl\samepage
Institut f\"ur Physik, Universit\"at Mainz, D-55099 Mainz, Germany$^{16}$}
\end{center}\end{sloppypar}
\vspace{2mm}
\begin{sloppypar}
\noindent
A.~Bonissent,
P.~Coyle,
C.~Curtil,
A.~Ealet,
D.~Fouchez,
P.~Payre,
A.~Tilquin
\nopagebreak
\begin{center}
\parbox{15.5cm}{\sl\samepage
Centre de Physique des Particules de Marseille, Univ M\'editerran\'ee,
IN$^{2}$P$^{3}$-CNRS, F-13288 Marseille, France}
\end{center}\end{sloppypar}
\vspace{2mm}
\begin{sloppypar}
\noindent
F.~Ragusa
\nopagebreak
\begin{center}
\parbox{15.5cm}{\sl\samepage
Dipartimento di Fisica, Universit\`a di Milano e INFN Sezione di
Milano, I-20133 Milano, Italy.}
\end{center}\end{sloppypar}
\vspace{2mm}
\begin{sloppypar}
\noindent
A.~David,
H.~Dietl,
G.~Ganis,$^{27}$
K.~H\"uttmann,
G.~L\"utjens,
W.~M\"anner,
\mbox{H.-G.~Moser},
R.~Settles,
G.~Wolf
\nopagebreak
\begin{center}
\parbox{15.5cm}{\sl\samepage
Max-Planck-Institut f\"ur Physik, Werner-Heisenberg-Institut,
D-80805 M\"unchen, Germany\footnotemark[16]}
\end{center}\end{sloppypar}
\vspace{2mm}
\begin{sloppypar}
\noindent
J.~Boucrot,
O.~Callot,
M.~Davier,
L.~Duflot,
\mbox{J.-F.~Grivaz},
Ph.~Heusse,
A.~Jacholkowska,$^{32}$
L.~Serin,
\mbox{J.-J.~Veillet},
J.-B.~de~Vivie~de~R\'egie,$^{28}$
C.~Yuan
\nopagebreak
\begin{center}
\parbox{15.5cm}{\sl\samepage
Laboratoire de l'Acc\'el\'erateur Lin\'eaire, Universit\'e de Paris-Sud,
IN$^{2}$P$^{3}$-CNRS, F-91898 Orsay Cedex, France}
\end{center}\end{sloppypar}
\vspace{2mm}
\begin{sloppypar}
\noindent
G.~Bagliesi,
T.~Boccali,
L.~Fo\`a,
A.~Giammanco,
A.~Giassi,
F.~Ligabue,
A.~Messineo,
F.~Palla,
G.~Sanguinetti,
A.~Sciab\`a,
R.~Tenchini,$^{1}$
A.~Venturi,$^{1}$
P.G.~Verdini
\samepage
\begin{center}
\parbox{15.5cm}{\sl\samepage
Dipartimento di Fisica dell'Universit\`a, INFN Sezione di Pisa,
e Scuola Normale Superiore, I-56010 Pisa, Italy}
\end{center}\end{sloppypar}
\vspace{2mm}
\begin{sloppypar}
\noindent
O.~Awunor,
G.A.~Blair,
G.~Cowan,
A.~Garcia-Bellido,
M.G.~Green,
L.T.~Jones,
T.~Medcalf,
A.~Misiejuk,
J.A.~Strong,
P.~Teixeira-Dias
\nopagebreak
\begin{center}
\parbox{15.5cm}{\sl\samepage
Department of Physics, Royal Holloway \& Bedford New College,
University of London, Egham, Surrey TW20 OEX, United Kingdom$^{10}$}
\end{center}\end{sloppypar}
\vspace{2mm}
\begin{sloppypar}
\noindent
R.W.~Clifft,
T.R.~Edgecock,
P.R.~Norton,
I.R.~Tomalin
\nopagebreak
\begin{center}
\parbox{15.5cm}{\sl\samepage
Particle Physics Dept., Rutherford Appleton Laboratory,
Chilton, Didcot, Oxon OX11 OQX, United Kingdom$^{10}$}
\end{center}\end{sloppypar}
\vspace{2mm}
\begin{sloppypar}
\noindent
\mbox{B.~Bloch-Devaux},
D.~Boumediene,
P.~Colas,
B.~Fabbro,
E.~Lan\c{c}on,
\mbox{M.-C.~Lemaire},
E.~Locci,
P.~Perez,
J.~Rander,
B.~Tuchming,
B.~Vallage
\nopagebreak
\begin{center}
\parbox{15.5cm}{\sl\samepage
CEA, DAPNIA/Service de Physique des Particules,
CE-Saclay, F-91191 Gif-sur-Yvette Cedex, France$^{17}$}
\end{center}\end{sloppypar}
\vspace{2mm}
\begin{sloppypar}
\noindent
N.~Konstantinidis,
A.M.~Litke,
G.~Taylor
\nopagebreak
\begin{center}
\parbox{15.5cm}{\sl\samepage
Institute for Particle Physics, University of California at
Santa Cruz, Santa Cruz, CA 95064, USA$^{22}$}
\end{center}\end{sloppypar}
\vspace{2mm}
\begin{sloppypar}
\noindent
C.N.~Booth,
S.~Cartwright,
F.~Combley,$^{31}$
P.N.~Hodgson,
M.~Lehto,
L.F.~Thompson
\nopagebreak
\begin{center}
\parbox{15.5cm}{\sl\samepage
Department of Physics, University of Sheffield, Sheffield S3 7RH,
United Kingdom$^{10}$}
\end{center}\end{sloppypar}
\vspace{2mm}
\begin{sloppypar}
\noindent
A.~B\"ohrer,
S.~Brandt,
C.~Grupen,
J.~Hess,
A.~Ngac,
G.~Prange,
U.~Sieler
\nopagebreak
\begin{center}
\parbox{15.5cm}{\sl\samepage
Fachbereich Physik, Universit\"at Siegen, D-57068 Siegen, Germany$^{16}$}
\end{center}\end{sloppypar}
\vspace{2mm}
\begin{sloppypar}
\noindent
C.~Borean,
G.~Giannini
\nopagebreak
\begin{center}
\parbox{15.5cm}{\sl\samepage
Dipartimento di Fisica, Universit\`a di Trieste e INFN Sezione di Trieste,
I-34127 Trieste, Italy}
\end{center}\end{sloppypar}
\vspace{2mm}
\begin{sloppypar}
\noindent
H.~He,
J.~Putz,
J.~Rothberg
\nopagebreak
\begin{center}
\parbox{15.5cm}{\sl\samepage
Experimental Elementary Particle Physics, University of Washington, Seattle,
WA 98195 U.S.A.}
\end{center}\end{sloppypar}
\vspace{2mm}
\begin{sloppypar}
\noindent
S.R.~Armstrong,
K.~Berkelman,
K.~Cranmer,
D.P.S.~Ferguson,
Y.~Gao,$^{29}$
S.~Gonz\'{a}lez,
O.J.~Hayes,
H.~Hu,
S.~Jin,
J.~Kile,
P.A.~McNamara III,
J.~Nielsen,
Y.B.~Pan,
\mbox{J.H.~von~Wimmersperg-Toeller}, 
W.~Wiedenmann,
J.~Wu,
Sau~Lan~Wu,
X.~Wu,
G.~Zobernig
\nopagebreak
\begin{center}
\parbox{15.5cm}{\sl\samepage
Department of Physics, University of Wisconsin, Madison, WI 53706,
USA$^{11}$}
\end{center}\end{sloppypar}
\vspace{2mm}
\begin{sloppypar}
\noindent
G.~Dissertori
\nopagebreak
\begin{center}
\parbox{15.5cm}{\sl\samepage
Institute for Particle Physics, ETH H\"onggerberg, 8093 Z\"urich,
Switzerland.}
\end{center}\end{sloppypar}
}
\footnotetext[1]{Also at CERN, 1211 Geneva 23, Switzerland.}
\footnotetext[2]{Now at Fermilab, PO Box 500, MS 352, Batavia, IL 60510, USA}
\footnotetext[3]{Also at Dipartimento di Fisica di Catania and INFN Sezione di
 Catania, 95129 Catania, Italy.}
\footnotetext[4]{Now at University of Florida, Department of Physics, Gainesville, Florida 32611-8440, USA}
\footnotetext[5]{Also Istituto di Cosmo-Geofisica del C.N.R., Torino,
Italy.}
\footnotetext[6]{Now at Institut f\"ur Experimentelle Kernphysik, Universit\"at Karlsruhe, 76128 Karlsruhe, Germany.}
\footnotetext[7]{Supported by CICYT, Spain.}
\footnotetext[8]{Supported by the National Science Foundation of China.}
\footnotetext[9]{Supported by the Danish Natural Science Research Council.}
\footnotetext[10]{Supported by the UK Particle Physics and Astronomy Research
Council.}
\footnotetext[11]{Supported by the US Department of Energy, grant
DE-FG0295-ER40896.}
\footnotetext[12]{Now at Departement de Physique Corpusculaire, Universit\'e de
Gen\`eve, 1211 Gen\`eve 4, Switzerland.}
\footnotetext[13]{Supported by the Commission of the European Communities,
contract ERBFMBICT982874.}
\footnotetext[14]{Supported by the Leverhulme Trust.}
\footnotetext[15]{Permanent address: Universitat de Barcelona, 08208 Barcelona,
Spain.}
\footnotetext[16]{Supported by Bundesministerium f\"ur Bildung
und Forschung, Germany.}
\footnotetext[17]{Supported by the Direction des Sciences de la
Mati\`ere, C.E.A.}
\footnotetext[18]{Supported by the Austrian Ministry for Science and Transport.}
\footnotetext[19]{Now at SAP AG, 69185 Walldorf, Germany}
\footnotetext[20]{Now at Groupe d' Astroparticules de Montpellier, Universit\'e de Montpellier II, 34095 Montpellier, France.}
\footnotetext[21]{Now at BNP Paribas, 60325 Frankfurt am Mainz, Germany}
\footnotetext[22]{Supported by the US Department of Energy,
grant DE-FG03-92ER40689.}
\footnotetext[23]{Now at Institut Inter-universitaire des hautes Energies (IIHE), CP 230, Universit\'{e} Libre de Bruxelles, 1050 Bruxelles, Belgique}
\footnotetext[24]{Also at Dipartimento di Fisica e Tecnologie Relative, Universit\`a di Palermo, Palermo, Italy.}
\footnotetext[25]{Now at McKinsey and Compagny, Avenue Louis Casal 18, 1203 Geneva, Switzerland.}
\footnotetext[26]{Now at Honeywell, Phoenix AZ, U.S.A.}
\footnotetext[27]{Now at INFN Sezione di Roma II, Dipartimento di Fisica, Universit\`a di Roma Tor Vergata, 00133 Roma, Italy.}
\footnotetext[28]{Now at Centre de Physique des Particules de Marseille, Univ M\'editerran\'ee, F-13288 Marseille, France.}
\footnotetext[29]{Also at Department of Physics, Tsinghua University, Beijing, The People's Republic of China.}
\footnotetext[30]{Now at SLAC, Stanford, CA 94309, U.S.A.}
\footnotetext[31]{Deceased.}
\footnotetext[32]{Also at Groupe d' Astroparticules de Montpellier, Universit\'e de Montpellier II, 34095 Montpellier, France.}  
\setlength{\parskip}{\saveparskip}
\setlength{\textheight}{\savetextheight}
\setlength{\topmargin}{\savetopmargin}
\setlength{\textwidth}{\savetextwidth}
\setlength{\oddsidemargin}{\saveoddsidemargin}
\setlength{\topsep}{\savetopsep}
\normalsize
\newpage
\pagestyle{plain}
\setcounter{page}{1}